\documentclass[11pt]{article}

\usepackage[hyphens]{url}
\usepackage{times}
\usepackage{ifthen}

\newboolean{lncs}
\setboolean{lncs}{false}   

\usepackage{pslatex}
\usepackage{amssymb}
\usepackage{latexsym}
\usepackage{graphicx}
\graphicspath{{../viewsize-data/}} 

\newcommand{\qxor}{\oplus}  

\makeatletter

\usepackage[normalem]{ulem}
 \usepackage{verbatim}

\ifthenelse{\boolean{lncs}}{}{
\newtheorem{theorem}{Theorem}[section]   
\newtheorem{lemma}{Lemma}    
 \newtheorem{proposition}{Proposition}  
 \newtheorem{corollary}{Corollary}  
 \newenvironment{proof}{\paragraph{Proof:}}{\hfill \ensuremath{\Box}}
}
\usepackage{setspace}
\usepackage{pslatex}
\newcommand{\komment}[1]{}

\usepackage{float}
\newboolean{hush}  \setboolean{hush}{true}
\ifthenelse{\boolean{hush}}{
\newcommand{\daniel}[1]{}%
\newcommand{\owen}[1]{}%
}{
\newcommand{\daniel}[1]{\textbf{DL: #1}}
\newcommand{\owen}[1]{\textbf{OK: #1}}
}

\newcommand{\notconference}[1]{#1}
\newcommand{\justconference}[1]{}

\usepackage{subfigure}

\begin{document}

\title{One-Pass, One-Hash $n$-Gram Statistics Estimation\footnote{
This is an expanded version of~\cite{lemi:one-pass-conference}.
}\\{\small \singlespacing Department of CSAS, UNBSJ\\ \vspace*{-1em} 
 Technical Report TR-06-001}}


  \ifthenelse{\boolean{lncs}}{
  \author{Daniel Lemire\inst{1} \and Owen Kaser\inst{2}}
  \institute{
  LICEF, T\'eluq\\
  Universit\'e du Qu\'ebec \`a Montr\'eal\\ 
  Montr\'eal, QC Canada\\
  \texttt{lemire@acm.org}
  \and
  Computer Science and Applied Statistics\\
  University of New Brunswick\\
  Saint John, NB Canada\\
  \texttt{o.kaser@computer.org}
} 
}{
   \author{
   Daniel Lemire\\
   Universit\'e du Qu\'ebec \`a Montr\'eal\\ 
   100 Sherbrooke West\\
   Montr\'eal, QC Canada\\
   lemire@acm.org\\
   \and
   Owen Kaser\\
   University of New Brunswick\\
   CSAS Dept.\\ Saint John, NB Canada\\ 
   o.kaser@computer.org\\
   }
   }
%
%

\maketitle

\begin{abstract}
In multimedia, text or bioinformatics databases,  applications
query sequences of $n$ consecutive symbols called $n$-grams.
Estimating the number of distinct $n$-grams is a view-size estimation
problem. While view sizes can be estimated by sampling
under statistical assumptions,
we desire an unassuming algorithm with universally valid accuracy
bounds. Most related work has focused on repeatedly hashing the
data, which is prohibitive for large data sources.
We prove that a one-pass one-hash algorithm is sufficient
for accurate estimates if the hashing is sufficiently independent.
To reduce costs further, we investigate recursive random hashing 
algorithms and show that they are sufficiently independent in practice. 
We compare our running times with exact counts using suffix arrays
and show that, while we use hardly any storage, we are an order
of magnitude faster. 
The approach further is extended to a one-pass/one-hash computation of $n$-gram
entropy and iceberg counts. 
The experiments use a large collection of
English text from the Gutenberg Project as well as synthetic data.
\end{abstract}

\komment{
\section{Notation}
{\footnotesize
$N$ stands for the number of symbols 

$n$ is the length of the $n$-grams

$m$ is the number of distinct symbols

$M$ is the available memory

 $I$ be the entire set of $n$-grams ($N=\textrm{card}(I)$)

$I'$ be the set of \textbf{probed} $n$-grams ($\textrm{card}(I')\leq M$)

$m'=\textrm{card}(I')$ ($m' \in [M/2 , M)$)

$f_i$ is the number of times the $n$-gram $i$ occurs ($\sum_i f_i= N$)

\texttt{a} and \texttt{b} are two arbitrary distinct
symbols.

Usual formal-language exponentiation as repetition.

Lambda calculus notation?

$p(x)$ and $q(x)$ are polys.  $p$ is degree of independence. It is also an arbitrary prime (eg, for MOD in a hash fcn). P(i) might be the probability
of an event $i$ or the probability of seeing a symbol $i$.
Owen does not like the overload.

$x_i$ is the $i^\textrm{th}$ stream item.

$\Sigma$ is the alphabet of symbols.
}

}

\section{Introduction}

Consider a sequence of symbols $a_i\in \Sigma$ of length $N$.~
\notconference{\komment{Some of the symbols are not
of interest such as ``white space'' or ``punctuation'' characters.}%
Perhaps the data source has high latency, for example, it is not in a flat
binary format or in a DBMS, making random access and skipping
impractical.}
 The symbols need not be characters from a natural
language: they can be particular ``events'' inferred from a sensor or
a news feed, they can be financial or biomedical patterns found in time
series, they can be words in a natural language, and so on. 
While small compared to the amount of memory available, the number
of distinct symbols ($\vert \Sigma \vert$) could be large:
on the order of $10^5$ in the case of words in a typical
English dictionary or $10^7$ in the case of the Google 5-gram data set
~\cite{goog:all-our-ngram}.
We make no other assumption
about the distribution of these distinct symbols.

An $n$-gram is a consecutive sequence of $n$ symbols. Given a data
source containing $N$ symbols, there are up to $N-n$ distinct
$n$-grams. \justconference{A multitude of applications exist for $n$-grams; see~\cite{cohenhash} for an overview. }%
\notconference{We use $n$-grams in language modeling~\cite{GaoZhang},
pattern recognition~\cite{Yannakoudakis1990}, predicting web page accesses~\cite{deshpande2004smm},
information
retrieval~\cite{Nie2000}, 
text categorization and author attribution~\cite{caropreso2001lie,joul:llc-attribution06,kese:CNG-method-entry}, speech recognition~\cite{jelinek1998sms},
multimedia~\cite{Paulus2003}, music retrieval~\cite{doraisamy2003pia},
text~mining~\cite{Losiewicz2000}, information
theory~\cite{Shannon1948}, software fault diagnosis~\cite{jagadeesh2005dma}, data compression~\cite{TelcorPatent}, 
%
%
%
%
%
%
data mining~\cite{Su2000},
indexing~\cite{KimWhang}, On-line Analytical Processing
(OLAP)~\cite{KeithKaserLemireAPICS2005}, optimal character recognition
(OCR)~\cite{droettboom2003cbc}, automated translation~\cite{1073465},
time series segmentation~\cite{cohen2002usc},
and so on.
}%
This paper concerns the use of  previously published hash functions for
$n$-grams, together with recent randomized algorithms for estimating
the number of distinct items in a stream of data.   Together, they
permit memory-efficient estimation of the number of distinct $n$-grams.

The number of distinct $n$-grams grows  large with $n$: 
Google makes available $1.1 \times 10^9$~word 5-grams, each occurring more than 400~times
 in $10^{12}$~words of text~\cite{goog:all-our-ngram}.  On a
smaller scale, 
storing
Shakespeare's First Folio~\cite{Gutenberg} takes up about 4.6\,MiB but
we can verify that it has over 3~million \textbf{distinct} 15-grams of
characters. If each distinct $n$-gram can be stored using 
$\log (4.6\times 10^6)\approx 22$~bits,
then we need about 8.4\,MiB just to store the $n$-grams 
\notconference{($3\times 10^6 \times 22 /8 \approx  8.3\times 10^6$) }%
without counting 
the indexing overhead.  
Thus, storing and indexing $n$-grams can use up more
storage than the original data source. 
Extrapolating this to the
large corpora used in computational linguistic studies, we see
the futility of using brute-force approaches that store the $n$-grams 
in main memory, when $n$ is large.  For smaller values of $n$, 
 $n$-grams of English words are also likely to defeat brute-force approaches.

\notconference{
Even if storage is not an
issue, indexing a very large number $m$ of distinct $n$-grams
 is computationally expensive because 
of the overhead associated with indexing data structures.
In practice, we  have a bound on
processing time or storage and we may need to 
focus on selected $n$-grams such as $n$-grams occurring more
than once or less than 10~times (iceberg $n$-grams)
\daniel{In~\cite{304214}, the case <K is specifically discussed (see page 7).
They say that while the Bottom-Up Cube algorithm fails, the other 
algorithms can still be used. So, while there might be further research
to be done on arbitrary predicates, considering arbitrary predicates is not
novel. In fact, recall that it is partially motivated by the HAVING 
clause in SQL.}%
depending on the
estimates.  Moreover, when only a count or an entropy estimate is
required for cost optimizers or user feedback, materializing all
$n$-grams is unnecessary. Finally, determining the number of infrequent
$n$-grams is important for building ``next word''
indexes~\cite{564415,ChangPoon2006} since inverted
indexes~\cite{Zobel1998} are most efficient over rare words or
phrases. Therefore, estimating online and quickly the $n$-gram statistics in a
single pass  is an important problem.
}%

\owen{made following wordier to fix linebreak problem.}%
There are two strategies for estimation of statistics of a sequence in a
single pass~\cite{Kearns1994}\notconference{\cite{Batu2002,Guha2006}}. 
The generative (or
black-box) strategy samples values at random.  From the samples, the
probabilities of each value is estimated by maximum likelihood or
other statistical techniques.  The evaluative strategy, on the other
hand, probes the exact probabilities or, equivalently, the number of
occurrences of (possibly randomly) chosen values. In
one pass, we can randomly probe several $n$-grams so we know their exact
frequency. 

On the one hand, it is difficult to estimate the number of distinct
elements from a sampling, without making further assumptions.  For
example, suppose there is only one distinct $n$-gram in 100~samples
out of 100,000~$n$-grams. Should we conclude that there is only
one distinct $n$-gram overall? Perhaps there are 100~distinct $n$-grams,
but 99 of them only occur once\notconference{, thus there is  a $\approx 91$\%~probability that we
observe only the common one}.  \notconference{While this example is a bit extreme,
skewed distributions are quite common as the Zipf law shows.
Choosing, a priori, the number of samples we require is a major
difficulty. Estimating the probabilities from sampling is a problem
that still interests researchers to this
day~\notconference{\cite{755182}}\cite{Orlitsky2003}.  }%

On the other hand, distinct count
estimates from a probing are statistically
easier~\cite{Gibbons2001}. With the example above, with just enough
storage budget to store 100~distinct $n$-grams, we would get an exact
count estimate!  On the downside, probing requires properly randomized
hashing.

In the spirit of probing, Gibbons-Tirthapura (GT)~\cite{Gibbons2001} count 
estimation goes as
follows.  We have $m$ distinct items in a stream containing the
distinct items $x_1,\ldots,x_m$ with possible repetitions. Let $h(x_i)$
be pairwise independent hash values over $[0,2^L)$ and let $h_t(x_i)$
be the first $t$ bits of the hash value. We have that $E(\textrm{card}
(\{h_t^{-1}(0)\}))=m/2^t$.  Given a fixed memory budget $M$, and
setting $t=0$, as we scan, we store all distinct items $x_i$ such that
$h_t(x_i)=0$ in a look-up table $H$. As soon as $\textrm{size}(H)=M+1$,
we increment $t$ by 1 and remove all $x_i$ in $H$ such that $h_t(x_i)
\neq 0$. Typically, at least one element in $H$ is removed, but
if not, the process of incrementing $t$ and removing items is
repeated until $\textrm{size}(H)<M$.  Then  we continue
scanning. After the run is completed, we return $\textrm{size}(H)
\times 2^t$ as the estimate.  By choosing
$M=576/\epsilon^2$~\cite{BarYossef2002}, we achieve an accuracy of
$\epsilon$, 5 times out of 6 ($P(|\textrm{size}(H) \times 2^t
-m|>\epsilon m) < 1/6$), by an application of Chebyshev's inequality.
By Chernoff's bound, running the algorithm $O(\log 1/\delta)$ times
and taking the median of the results gives a reliability of $\delta$
instead of 5/6. Bar-Yossef {et al.} suggest to improve the  
algorithm by storing hash values of the $x_i$'s instead of the $x_i$'s
themselves, reducing the reliability but lowering the memory usage.
Notice that our Corollary~\ref{antibar} shows that the estimate of
a 5/6 reliability for $M=576/\epsilon^2$ is pessimistic: 
$M=576/\epsilon^2$
implies a reliability of over 99\%. We also prove that replacing pairwise
independent by 4-wise independent hashing substantially improves the existing
theoretical performance bounds\footnote
{  
The application of $p$-wise independent hash functions for estimating
frequency moments is well known~\cite{bhuvanagiri2006sae, indyk2005oaf}.
}.

Random hashing can be the real bottleneck in probing, but to alleviate
this problem for $n$-gram hashing, we use recursive
hashing~\cite{cohenhash,karp1987erp}: we  leverage the fact
that successive $n$-grams have $n-1$ characters in common.  We study
empirically online $n$-gram 
\notconference{statistical estimations (counts, iceberg counts
and entropy)}%
\justconference{count estimation}
in one pass that hashes each $n$-gram only once. 
We compare several different recursive $n$-gram hashing algorithms 
including hashing by cyclic and irreducible polynomials in the binary
Galois Field ($\textrm{GF}(2)[x]$). 
The main contributions of this paper are a tighter theoretical bound
in count estimation and an experimental validation to demonstrate
practical usefulness.
This work has a wide range of applications, from other
view-size estimation problems to text mining.

\section{Related Work}
\label{relatedwork}

\notconference{Related work includes reservoir sampling, suffix arrays, and 
view-size estimation in OLAP.}%

\notconference{
\subsection{Reservoir Sampling}
We can choose randomly, without replacement, $k$ samples in a
sequence of unknown length using a single pass through the data by
\emph{reservoir sampling}. Reservoir 
sampling~\cite{3165}\notconference{\cite{Kolonko2006SRS,198435}}
 was introduced by
Knuth~\cite{Knuth1969}.  All reservoir sampling algorithms begin by
appending the first $k$ samples to an array. In their linear time
($O(N)$) form, reservoir sampling algorithms sequentially visit every
symbol choosing it as a possible sample with probability $k/t$ where
$t$ is the number of symbols read so far. 
\notconference{The chosen sample is simply
appended at the end of the array while an existing sample is flagged
as having been removed. The array has an average size of $k \left (
1+\log N/k \right )$ samples at the end of the run. In their sublinear
form ($O(k(1 + \log(N/k))$ expected time), the algorithms skip a random
number of data points each time. }%
While these algorithms use a single
pass, they assume that the number of required samples $k$ is known a
priori, but this is difficult without any knowledge of the data
distribution.}%

\notconference{\subsection{Suffix Arrays}}%
Using suffix arrays~\notconference{\cite{320218}}\cite{suffix-array-journal} 
and the length
of the maximal common prefix between successive prefixes, Nagao and
Mori~\cite{NagaoMori} proposed a fast algorithm to compute
$n$-gram statistics exactly. However, it cannot be considered an
 online algorithm even if we compute the suffix array in one pass:
after constructing the suffix array, one must
go through all suffixes at least once more.
 Their implementation was later improved by Kit
and Wilks~\cite{kit1998vca}. \notconference{Unlike suffix
trees~\cite{lazysuffixtrees}, uncompressed suffix arrays do not
require several times the storage of the original document and their
performance does not depend on the size of the alphabet.}
Suffix arrays
can be constructed in $O(N)$ time using $O(N)$ working
space~\cite{HonSung}. Querying a suffix array for a given $n$-gram
takes $O(\log N)$ time.

\notconference{\subsection{View-Size Estimation in OLAP}}
By definition, each $n$-gram is a tuple of length $n$ and can be
viewed as a relation to be aggregated. OLAP (On-Line
Analytical Processing)\notconference{~\cite{codd93}} is a
database acceleration technique used for deductive analysis, typically
involving aggregation.  To
achieve acceleration, one frequently builds data
cubes\notconference{~\cite{graycube}}
where multidimensional relations are pre-aggregated
in multidimensional arrays. 
\notconference{OLAP is commonly used
for business purposes with dimensions such as time, location, sales, expenses, and so on.
Concerning text, most work has
focused on informetrics/bibliomining, document management and
information
retrieval~\cite{345656,moth:doccube,niem:mddata-model-informetrics,bern:a-juste-titre,Sullivan2001}.
The idea of using OLAP for exploring the text content itself
(including phrases and $n$-grams) was proposed for the first time by
Keith, Kaser and Lemire~\cite{KeithKaserLemireTR05001,KeithKaserLemireAPICS2005}.
}
The estimation of $n$-gram counts can be viewed as an OLAP
view-size estimation problem which itself ``remains an important area
of open research''~\cite{dehne2006cpo}.  A data-agnostic approach to
view-size estimation~\cite{shukla:sem}, which is likely to be used by
database vendors, can be computed almost instantly\justconference{.}
\notconference{ as long as we know
how many attributes each dimension has and the number of relations
$\eta$.  For $n$-gram
estimation, the number of attributes is the size of the alphabet $\vert \Sigma \vert$
and $\eta$ is the number of $n$-grams with possible repetitions ($\eta=N-n+1$).
\komment{
\daniel{$\eta$ is known ahead of time, and so is $V$. I think your
problem was my notation, which was a bit confusing.}
\owen{will check later.  Since this technique has not been proposed for ngrams, is it
worthwhile see how it predicts, say, 5-grams based on 15-grams, over Gutenberg inputs?}
\daniel{It might be interesting to include it in the experimental section. 
However, it is simpler than you seem to think. Let $K$ be the number of 
distinct characters, say $K=35$, then $V=35^n$ and the number of
$n$-gram is approximated by $(1-\frac{1}{35^n})^{N-n+1} \times
(N-n+1)$ where $N$ is the length of the text.}
\owen{Are you sure that this is their model? Or at least, this is how
they are applying it?  I have been trying to
mentally relate this to data-agnostic models that Richard and I used
and that were used in the cube-compress tech report (I think).  The
Panda people would (I think) be doing something similar since they're 
all data cube
people too.  In that world, you're asking, if you drop dimension 1 in
a $d$ dimensional cube with uniform random density, ``what is the
probability that a group of $|D_1|$ cells all are unallocated'', since
that is the probability of unallocation of the $d-1$ dim cube.
Since this model is so crude, it is only semisensible when using it to
reduce by one or two dimensions---in our case, estimate 5-grams from the
number of 6-grams or 7-grams.   My prediction of 5-grams from 6-grams
is thus $N \times ( 1 - \delta^{35})$ where $N$ is the number of
6-grams and $\delta = N / 35^6$ is the density of 6-grams in their
cube.  Even if this turns out to be the same thing (not sure) it seems
more legit to predict  $n-1$-grams or $n-2$-grams from $n$-grams than
what I think you've said. }
}
}
 
Given $\eta$  cells picked uniformly at random, with replacement,
 in a $V=K_1\times K_2\times
\cdots K_n$ space, the probability that any given cell (think
``$n$-gram'') is omitted is $(1-\frac{1}{V})^{\eta}$. 
For $n$-grams, $V=\vert \Sigma \vert ^n$.
Therefore,  
the expected number of \textbf{un}occupied cells is $(1-\frac{1}{V})^\eta \times
\eta. $

Similarly, assuming the number of $n$-grams is known to be $m$, the same
model permits us to estimate the number of $n-1$-grams by
$m \times (1-(\frac{m}{V})^{|\Sigma|})$.
In practice,\notconference{these approaches overestimate}%
\justconference{this approach overestimates}%
 systematically because relations are not uniformly distributed.

A more sophisticated view-size estimation algorithm used in the
context of data warehousing and OLAP\notconference{~\cite{shukla:sem,Kotidis2002}}
is logarithmic probabilistic counting~\cite{flajolet1985pca}.  This
approach requires a single pass and almost no memory, but it assumes
independent hashing for  which 
no algorithm using limited storage is known~\cite{BarYossef2002}.
Practical results are sometimes
disappointing~\cite{dehne2006cpo}, possibly because 
many 
random hash values 
need to be computed for
each data point. \notconference{Other variants of this approach include
linear probabilistic counting~\cite{whang1990ltp,shah2004sem}
and loglog counting~\cite{durand2003lcl}.}

\notconference{
View-size estimation through sampling has been made adaptive by Haas
et al.~\cite{haas1995sbe}: their strategy is to first attempt
to determine whether the distribution is skewed and then use an appropriate
statistical estimator.  We can also count the marginal frequencies of
each attribute value (or symbol in an $n$-gram setting) and use them
to give estimates as well as (exact) lower and upper bound on the view
size~\cite{YuZuzarteSevcik}.  Other researchers make particular
assumptions on the distribution of 
relations~\cite{nadeau2003pmo,ciaccia2001eca,ciaccia2003bca,faloutsos1996msd}.
}

\section{Multidimensional Random Hashing}

Hashing encodes an object as a fixed-length bit string 
for comparison.
Multidimensional hashing is a particular form of hashing where the objects
can be represented as tuples. \notconference{Multidimensional hashing is
of general interest since several commonly occurring objects can be thought of
as tuples: 32-bit values can be seen as 8-tuples containing
4-bit values.}

\justconference{
Hashing is \textit{pairwise independent} or
\textit{universal}~\cite{carter1979uch} if $P(h(x_1)=y \land h(x_2)=z)=
P(h(x_1)=y) P(h(x_2)=z)=1/4^L$.
Gibbons and Tirthapura showed that
pairwise independence was sufficient to approximate count
statistics~\cite{Gibbons2001} essentially because the variance of the
sum of pairwise independent variables is just the sum the variances
($\textrm{Var}(X_1+\ldots+X_m)=\textrm{Var}(X_1)+\ldots+\textrm{Var}(X_m)
$). Pairwise independence implies uniformity.
An example of a pairwise independent hash function is $h(x)=a+bx \pmod{2^L}$
where $a$ and $b$ are chosen randomly.
}

For convenience, we consider hash functions mapping keys to
$[0,2^L)$, where the set $U$ of possible keys is much larger than $2^L$.
A difficulty with hashing is that any particular hash
function $h$ has some ``bad inputs'' $S \subset U$ over
which some hash value (such as 0) is either too frequent or not frequent enough
($\textrm{card}(h^{-1}(0)) \not \approx \textrm{card}(S)/2^L$) making
count estimates from hash values difficult. 
Rather than
make assumptions about the probabilities of bad inputs for a particular
fixed hash function, an alternative approach~\cite{carter1979uch}
selects the hash function randomly from some family $\mathcal{H}$ of
functions, all of which map $U$ to $[0,2^L)$.

Clearly, some families $\mathcal{H}$ have desirable properties that other
families do not have. 
\notconference{ For instance, consider a family whose members
always map to even numbers --- then considering the random possible
selections of $h$ from $\mathcal{H}$, for any $x \in U$ we
have $P(h(x)=i) = 0$ for any odd $i$.  This would be an undesirable property
for many applications.  We now mention some desirable properties of families.}
$\mathcal{H}$ is \textit{uniform} if,  considering $h$ selected
uniformly at random from $\mathcal{H}$ and for all $x$ and $y$, we have $P(h(x)=y)=1/2^L$. 
This condition is too weak; the family of constant functions is uniform but
would be disastrous when used with the GT algorithm.
We need stronger conditions implying that any particular 
member $h$ of the family must hash objects evenly over $[0,2^L)$.
 $\mathcal{H}$ is \textit{pairwise independent} or
\textit{universal}~\cite{carter1979uch} if 
for all $x_1$, $x_2$, $y$, $z$ with $x_1 \not = x_2$, we have that
$P(h(x_1)=y \land h(x_2)=z)= P(h(x_1)=y) P(h(x_2)=z)=1/4^L$.
We will refer to such an $h \in \mathcal{H}$ as a ``pairwise independent
hash function'' when the family in question can be inferred from the
context or is not important. Pairwise independence implies uniformity. 

Gibbons and Tirthapura showed that
pairwise independence was sufficient to approximate count
statistics~\cite{Gibbons2001} essentially because the variance of the
sum of pairwise independent variables is just the sum the variances
($\textrm{Var}(X_1+\ldots+X_j)=\textrm{Var}(X_1)+\ldots+\textrm{Var}(X_j)
$). \notconference{
A well-known example of a pairwise-independent hash function for keys in the range $[0,B^{r+1})$, where $B$ is prime, is computed as follows.
Express key $x$ as $x_r x_{r-1} \ldots x_0$ in base $B$.  Randomly
choose a number $a \in [0,2^{r+1})$ and express it as $a_r a_{r-1}\ldots a_0$
in base $B$.  Then,
set $h(x)=\sum_{i=0}^r a_i x_i \pmod{B}$. The
proof that it is pairwise independent follows from the fact that integers
 modulo a prime numbers form a field (GF($B$)).
}

Moreover, the idea of pairwise independence can be generalized:
a family of hash functions $\mathcal{H}$ is \emph{$k$-wise independent} 
if given distinct $x_1,\ldots,x_k$ and given $h$ selected
uniformly at random from $\mathcal{H}$,  then
$P(h(x_1)=y_1 \land \cdots \land h(x_k)=y_k)=1/2^{kL}$.  Note that $k$-wise
independence implies $k-1$-wise
independence and uniformity. 
(Fully) independent hash functions are $k$-wise independent for arbitrarily large
$k$.
\komment{
\textbf{DL: I don't have access to the original paper since UQAM is not
yet subscribed to IEEE. By reading related papers, I figured out that
$O(1)$ is in a RAM model, where the number of bits might be bounded
by the number of bits of the domain plus the number of bits of the range,
of the hashing function. I think the RAM model is a trick to hide 
the dependence on $k$. I've also changed the notation to remove the $p$.}}
Siegel~\cite{siegel1989ucf} has shown $k$-wise-independent
hash functions from $[0,2^k)$ to $[0,2^k)$ can be evaluated in $O(1)$ 
time in a unit-cost RAM model.  Whether his approach can be efficiently
implemented is unclear, and in any case his results are not directly 
applicable to us.  For instance, 
we assume a cost of $\Omega(n)$ to process an $n$-gram\footnote{
We \emph{do} assume O(1) time to access the first or last symbol
in an $n$-gram and O(1) expected time to find such a symbol in
a look-up table.  As well, we also implicitly make unit-cost 
assumptions when calculating the $L$-bit hash values.}.  
This paper contributes better bounds for approximate count
statistics, providing that more fully independent hash functions are
used (4-wise instead of pairwise, for instance).

\komment{
\owen{So far, this collision free and almost collision free stuff unused}
\daniel{Yes, can be removed.}
For practical purposes, it is sometimes useful to use a
hash function $h$ where collisions are unlikely.  Using $h$, we can then count
accurately the number of distinct hash values instead of counting the
number of distinct tuples.  The probability of a collision for a
pairwise independent hash function is $P(h(x_1)=h(x_2))=1/2^L$.
If we consider
$m$ different values $x_1, x_2, \ldots, x_m$ hashed by $h$ then
the expected number of collisions is ${m\choose 2} \times
\frac{1}{2^L}= \frac{m(m-1)}{2^{L+1}}\approx \frac{m ^2}{2^{L+1}}$.  A
hash function is \emph{almost collision-free} if the expected number of
collisions is less than 1, which is true if $m < 2^{(L+1)/2}$. Choosing $L=64$ means that the hash is almost
collision-free as long as $m$ is smaller than $6 \times 10^9$.
}

In the context of $n$-gram hashing, we seek  \textit{recursive} families of
hash functions
so that we can compute new hash values
quickly by reusing previous hash values. A hash function $h$ is recursive
if there exist a (fixed) function $F$ over triples such that
\[h(x_2,x_3,\ldots,x_{n+1})= F( h(x_1,x_2,\ldots,x_n) , x_1 , x_{n+1}).\]
The function $F$ must be independent of $h$. ($F$ is common to all members of the family).
By extension\footnote
{ This extended sense resembles Cohen's use of ``recursive.''
}, a hash function $h$ is 
\emph{recursive over  hash values} $\tau(x_i)$, where $\tau$ is
a randomized hash function for symbols,
if there is a function $F$, independent of $\tau$ and $h$ 
such that 
\[h(x_2,\ldots,x_{n+1})=F(h(x_1,\ldots,x_n),\tau(x_1),\tau(x_{n+1})).\]
\owen{``randomized hash function'' is not the usual family-oriented wording.}

Similarly, a hash function $h$ is \emph{semi-recursive} if there exists a
(fixed) function $G$ over pairs such that
\[h(x_1,x_2,\ldots,x_{n+1})= G( h(x_1,x_2,\ldots,x_n) ,x_{n+1}).\]
This relates hashing an $n+1$-gram to hashing an overlapping
$n$-gram, whereas recursive hashing involves two overlapping $n$-grams.
Hence, we cannot say that  recursive hash functions are
semi-recursive.

As an example of a recursive hash function,
given tuples  $(x_1,x_2,\ldots,x_n)$ whose components are integers taken from $[0,B)$, 
we can hash  by
the Karp-Rabin formula $\sum_i x_i B^{i-1} \bmod{R}$, where $R$ is some
prime defining the range of the hash
function~\cite{karp1987erp,gonnet1990akr}. 
This is \notconference{semi-recursive, 
with $G(v,x_{n+1}) = (v + x_{n+1} B^n) \bmod{R}$ 
and also}
recursive.
Regardless, it is a poor hash function for us, since
$n$-grams with common
suffixes all get very similar hash values.  
For probabilistic-counting
approaches based on the number of trailing zeros of the hashed value, if
$h( x_1, x_2, \ldots, x_n)$ has many trailing zeros, then we know 
that $h( x_1, x_2, \ldots,x_{n-1}, x'_n)$  has few trailing
zeros (assuming $x_n \not = x'_n$).

In fact, no recursive hash function
can be pairwise independent. 

\begin{proposition}
 Recursive hash functions are not pairwise independent.
\end{proposition}
\begin{proof}
\notconference{
Suppose $h$ is a recursive pairwise independent hash function.
Let $h_{i,j}= h(x_i,\ldots,x_{j})$.
Then $F(h_{1,n},x_1,x_{n+1})=h_{2,n+1}$ where
the function $F$ must be independent of $h$.
\komment{\daniel{The trick here is that $F$ is really not randomized in
any way. That's important otherwise any (randomized) hash function
is recursive. On the other hand, $h$ \textbf{is} picked randomly from a set
of hash functions (in Owen's linguo). However, recursive hash functions
are just not very random. To see how they behave, pick a de Bruijn sequence
(a sequence containing all $n$-grams). For a fixed $F$, start anywhere, and give a 
random hash value to the first $n$-gram you encounter, then all other $n$-gram
hash values are uniquely determined by running through the de Bruijn sequence!}
\owen{I disagree with the idea that ``any'' (or almost all) etc randomized hash
functions would otherwise be recursive.  If we assume that $U$ (the set of
potential keys) is much bigger than $2^L$, then potential collisions are
inevitable.  The ``fixup function'' $F$ therefore cannot be inverted to 
discover which $n$-gram produced it, (which I think is the basis for your
statement, because then you could shift in the new character, get a new $n$ gram
and then hash it to create $F$.)  I believe that you just don't have enough
information for an $F$ with only the shift-out character, the shift-in character.
Eg, suppose $h(x_1, x_2, \ldots , x_n) = h(x_1, x'_2, x'_3, x'_n)$
but $h(x_2, \ldots , x_n, x_{n+1}) \not = h(x_1, x'_2, x'_3, \ldots, x'_n, x_{n_1})$,
where $\neg \forall_{2 \leq i \leq n} x_i = x'_i$.  Then $F(h_{1,n}, x_1, x_n)$ cannot
be defined.  The suppositions, that two different ngrams starting with the same symbol might
collide (but their stream-successors won't collide) seems quite weak.}
               }

Fix the
values $x_1,\ldots, x_{n+1}$,
then 
\begin{eqnarray*}
\lefteqn{P(h_{1,n}=a, h_{2,n+1}=c)}&&\\
& = &P(h_{1,n}=a, F(a,x_1,x_{n+1})=c)\\
& = & \left \lbrace \begin{array}{ll}
P(h_{1,n}=a) & \mbox{if $F(a,x_1,x_{n+1})=c$}\\
0 & \mbox{otherwise}\end{array} \right . ,
\end{eqnarray*}
a contradiction.}
\justconference{See \cite{viewsizetechreport}.}
~\end{proof}

As the next lemma and proposition show, being recursive over 
hashed values, while a weaker requirement, does not allow
more than pairwise independence.

\begin{lemma}\label{bijectivelemma}
Let $F$ be the recursive function of any recursive uniform hash function,
then, for $v,w$ fixed, 
$\lambda x . F(x,v,w)$  
is one-to-one.
\end{lemma}

\begin{proof}
 We need to show that $F(x,v,w)=y$ and $F(x',v,w)=y$ implies $x=x'$.
Consider a sequence,
$v,x_2,\ldots,x_n,w$ and any uniform hash function $h$, then
\begin{eqnarray*}
\frac{1}{2^L}& = &P(h(x_2,\ldots,x_n,w)=y)  =  P(F(h(v,x_2,\ldots,x_n ),v,w)=y)\\
& = & \sum_{\eta \in \{z|F(z,v,w)=y\}} P(h(v,x_2,\ldots,x_n )=\eta) =  \sum_{\eta \in \{z|F(z,v,w)=y\}}\frac{1}{2^L}\\
& = & \frac{\textrm{card}(\{z|F(z,v,w)=y\})}{2^L},
\end{eqnarray*}
and hence $\textrm{card}(\{z|F(z,v,w)=y\})=1$ showing the result.~\end{proof}

\begin{proposition}\label{threewiseprop}
Recursive hashing functions over hashed values cannot be 3-wise independent.
\end{proposition}
\begin{proof}
Consider the string of symbols $\texttt{a}^n\texttt{bb}$, recalling
that $\texttt{a}$ and $\texttt{b}$ are arbitrary but 
distinct members of $\Sigma$.

We have that
\newcommand{\abfont}[1]{\texttt{#1}}
\begin{eqnarray*}
&P(h(\abfont{a},\ldots,\abfont{a})=x,h(\abfont{a},\ldots, \abfont{a},\abfont{b})=y,h(\abfont{a},\ldots, \abfont{a},\abfont{b},\abfont{b})=y)\\
&=  P(h(\abfont{a},\ldots,\abfont{a})=x,F(x,\tau(\abfont{a}),\tau(\abfont{b}))=y,F(y,\tau(\abfont{a}),\tau(\abfont{b}))=y).
\end{eqnarray*}
However, we can only have $F(x,\tau(\abfont{a}),\tau(\abfont{b}))=y$ and $F(y,\tau(\abfont{a}),\tau(\abfont{b}))=y$
if $x=y$ by Lemma~\ref{bijectivelemma} and so, the above probability is zero unless
$x=y$, preventing 3-wise independence.~\end{proof}

A trivial way to generate an independent hash is to  assign a
random integer in $[0,2^L)$ to each new value $x$. Unfortunately, this
requires as much processing and storage as a complete indexing of all
values.  However, in a multidimensional setting this approach can be
put to good use. Suppose that we have tuples in 
$K_1 \times K_2
\times \cdots \times K_n$ such that $\vert K_i \vert$ is small for
all $i$. We can  construct independent hash functions 
$h_i : K_i \rightarrow [0,2^L)$ for all $i$ and combine them. 
For example, the hash function
$h(x_1,x_2,\ldots,x_n)=h_1(x_1)\qxor h_2(x_2) \qxor \cdots \qxor
h_n(x_n)$ is $n$-wise independent ($\qxor$ is the exclusive or).
As long as the sets $K_i$,
are small,  in time $O(\sum_i \vert K_i \vert)$
we can construct the hash function by generating
$\sum_i \vert K_i \vert$ random numbers  and storing them in a look-up table.
With constant-time look-up, hashing an $n$-gram thus takes 
$O(L n)$ time, or $O(n)$ if $L$ is considered a constant.

\owen{I cannot think of any reason why they should not be stored in
arrays, for *fast* constant time look-up --- especially with $n$-gram
hashing, where you know that every $h_i$ gets to process every symbol
that occurs at least once in the text.  All I really need is some
mapping maps $\sigma_i$ to $i$ and that is pretty easy preprocessing.}
Unfortunately, this hash function is not recursive\justconference{.}
\notconference{but it is semi-recursive.  \owen{still not sure we want to talk about this for conference}%
}
In the
$n$-gram context, we can choose $h_1=h_2=\ldots$ since
$\Sigma=K_1=K_2=\ldots$  While the resulting hash function is recursive
over hashed values\notconference{ since 
\begin{eqnarray*}
h(x_2,\ldots,x_{n+1}) & = & h_1(x_2) \qxor \cdots \qxor h_1(x_{n+1})\\
& = & h_1(x_1) \qxor h_1(x_{n+1})\qxor h(x_1,\ldots,x_{n}),
\end{eqnarray*}}
\justconference{,}
it is
no longer even pairwise independent, since
$P(h(a,b)=w, h(b,a)=v)=0$ if $w\not =v$. 

To obtain some of the speed benefits of recursive hashing with $p\geq2$-wise
independence, a hybrid approach might be useful when $n$
is large.  It splits the $n$-gram into $p$ pieces, each of which can be
updated.  For simplicity, suppose $n$ is a multiple of $p$ and $n \geq 2p$.
We use
\begin{eqnarray*}
h(x_1,x_2,\ldots x_n) & = & h_1(x_1) \qxor h_1(x_2) \qxor 
    \ldots \qxor h_1(x_{n/p}) \\
&& \qxor h_2(x_{(n/p)+1}) \qxor \ldots \qxor h_2(x_{2n/p}) \\
&& \qxor \ldots\\
&& \qxor h_p(x_{((p-1)n/p)+1}) \qxor \ldots \qxor h_p(x_n).
\end{eqnarray*}
To update, note that 
\begin{eqnarray*}h(x_2,\ldots,x_{n+1}) & = &
h(x_1,\ldots,x_n) 
  \qxor h_1(x_1) \\  
&&\qxor (h_1(x_{n/p}+1) \qxor h_2(x_{n/p}+1)) \\ 
&&\qxor \ldots \\
&&\qxor (h_{p-1}(x_{((p-1)n/p)+1}) \qxor h_{p}(x_{((p-1)n/p)+1}) \\
&&\qxor h_p(x_{n+1}). 
\end{eqnarray*}
Thus, we have a $p$-wise independent recursive hash function.

\komment{
\textbf{Do we need to prove pwise indpendence or is it obvious enough?}  
\textbf{DL: I reserve the right to say I was wrong, but I think it is obvious.}
}

For $n$-gram estimation,
we seek families of hash functions that  
behave, for practical purposes, like $n$-wise independent 
while being recursive over hash values.
A particularly interesting form of hashing using the binary Galois 
field GF(2) is called ``Recursive Hashing by Polynomials'' and has been
attributed to Kubina by Cohen~\cite{cohenhash}.  
GF(2) 
contains
only two values (1 and 0)  with the addition (and hence subtraction)  
defined by ``exclusive or'', $a + b = a \qxor b$ and the
multiplication by ``and'', $a \times b = a \wedge b$. 
$\textrm{GF}(2)[x]$ is the
vector space of all polynomials with coefficients from GF(2).  Any
integer in binary form (e.g. $c=1101$) can thus be interpreted as an
element of $\textrm{GF}(2)[x]$ (e.g. $c=x^3+x^2+1$). If 
$p(x)\in \textrm{GF}(2)[x]$,
then $\textrm{GF}(2)[x]/p(x)$ can be thought of as $\textrm{GF}(2)[x]$ modulo
$p(x)$. 
As an example, if $p(x)=x^2$, then $\textrm{GF}(2)[x]/p(x)$ is the
set of all linear polynomials.  For instance, $x^3+ x^2+ x+1= x+1 \pmod{x^2}$ since,
in $\textrm{GF}(2)[x]$,
$(x+1) + x^2 (x+1) = x^3 + x^2 + x + 1$.

\owen{There is a problem.  For global consistency, $x_i$ is appropriate.
Yet I feel that having the dummy variable in the polynomial called ``x''
is not nice, as it seems to suggest that the $x_i$'s are parts of $x$.
}
\daniel{I don't follow, but I used Cohen's notation for this part of the paper.}
\owen{I think it is bad notation to have unrelating things sharing the
same ``base variable name''.  When I see an $x'$ or $\hat x$ or $x_i$ and there
is a plain $x$ floating around, I expect they ought to be related in some
interesting way. If I don't see it, I may think I am lost.}

Interpreting $h_1$ hash values as polynomials in $\textrm{GF}(2)[x]/p(x)$, 
and with the condition that $\textrm{degree}(p(x))\geq n$, we
define $h(a_1,a_2,\ldots,a_n)=h_1(a_1) x^{n-1} + h_1(a_2) x^{n-2}+
\ldots + h_1(a_n)$. 
It \emph{is} recursive over the sequence
$h_1(a_i)$. The combined
hash can be computed in constant time with respect to $n$ by reusing
previous hash values:
\[h(a_2,a_3,\ldots,a_{n+1})= x h(a_1,a_2,\ldots,a_n) - h_1(a_1) x^{n} + h_1(a_{n+1}).\]
\komment{\daniel{This is not true in general, but might true in particular cases:
A variant is pairwise independent and recursive (over hash values):
\[h(x_1,x_2,\ldots,x_n)=h_1(x_1) q(x)^{n-1} + h_1(x_2) q(x)^{n-2}+
\ldots + h_1(x_n)\]
where $q(x)$ is an element of $\textrm{GF}(2)[x]/p(x)$ chosen at random.}}
\komment{No recursive hash functions over hash values can be more than pairwise
independent (see Proposition~\ref{threewiseprop}).
We focus on the case $q(x)=x$ so that the computation of of $F$ can 
be done in time $O(L)$ as opposed to $O(L^2)$.}

Choosing $p(x)=x^{L}+1$ for $L\geq n$, for any polynomial 
$q(x) = \sum_{i=0}^{L-1} q_i x^i$, we have 
\[ x q(x) = x ( q_{L-1} x^{L-1}+  \ldots +q_1 x + q_0) = q_{L-2} x^{L-1}+ \ldots +  q_0 x + q_{L-1}.\]
Thus, we have that multiplication by $x$ is a cyclic left shift.
The resulting hash is called 
\textsc{Recursive Hashing by Cyclic Polynomials}~\cite{cohenhash}, or (for
short) \textsc{Cyclic}.
It was shown
\emph{empirically} to be uniform~\cite{cohenhash},
but it is not formally so:

\begin{lemma}\label{not-uniformlemma}
 \textsc{Cyclic} is not uniform for $n$ even and never pairwise independent.
\end{lemma}
\begin{proof}

If $n$ is even,  $x^{n-1}+\ldots+x+1$ is divisible by $x+1$, so
 $x^{n-1}+\ldots+x+1=(x+1)r(x)$ for some polynomial $r(x)$. Clearly,
$r(x) (x+1)(x^{L-1}+x^{L-2}+\ldots+x+1)=0  \pmod{x^L+1}$ for any $r(x)$ and so 
$P(h(\texttt{a}^n)=0)=P((x^{n-1}+\ldots+x+1)h_1(\texttt{a})=0)=
P((x+1) r(x) h_1(\texttt{a})=0)
\geq P(h_1(\texttt{a})=0 \lor h_1(\texttt{a})=r^{-1}(x) (x^{L-1}+x^{L-2}+\ldots+x+1))= 1/2^{L-1}$. Therefore,  \textsc{Cyclic} is not uniform for $n$ even.


To show \textsc{Cyclic} is never pairwise independent, consider $n=3$ (for simplicity), 
then $P(h(\texttt{aab})= h(\texttt{aba}))=P((x+1)(h_1(\texttt{a})+h_1(\texttt{b}))=0)\geq P(h_1(\texttt{a})+h_1(\texttt{b})=0 \lor h_1(\texttt{a})+h_1(\texttt{b})=x^{L-1}+x^{L-2}+\ldots+x+1)=1/2^{L-1}$, but pairwise independent
hash values are equal with probability $1/2^L$. The result is shown.
\end{proof}

In contrast to \textsc{Cyclic},
to generate hash functions over
$[0,2^L)$ we can choose $p(x)$ to be an irreducible polynomial 
of degree $L$ in
$\textrm{GF}(2)[x]$.  For $L=19$, an example choice is $p(x)= 1 +
x^2+x^3+x^5+x^6+x^7+x^{12}+x^{16}+x^{17}+x^{18}+x^{19}$~\cite{COS}.
\notconference{
(With this particular irreducible 
polynomial, $L=19$ and so we require $n \leq 19$.  Irreducible polynomials of
larger degree can be found \cite{COS} if desired.)  
Computing $(a_{18} x^{18}+\ldots +a_0 ) x \pmod {p(x)}$
as a polynomial of degree 18 or less, for representation in 
19 \owen{was 18, I changed but am not sure} bits, can be done efficiently.
We have 
$(a_{18} x^{18}+\ldots +a_0 ) x  = a_{18}(p(x)-x^{19})+a_{17}x^{18}\ldots +a_0 x \pmod {p(x)}$
and the polynomial on the right-hand-side is of degree at most 18. In practice, we do a left shift of the first
18 bits of the hash value and if the value of the 19$^\textrm{th}$ bit
is 1, then we apply an exclusive or with the integer $1 +
2^2+2^3+2^5+2^6+2^7+2^{12}+2^{16}+2^{17}+2^{18}+2^{19}$.  The
resulting hash is called \textsc{Recursive Hashing by General
Polynomials}~\cite{cohenhash}, or (for short) \textsc{General}.
The main benefit of setting $p(x)$ to
be an irreducible polynomial is that $\textrm{GF}(2)[x]/p(x)$ is a field; in
particular, it is no longer possible that $p_1(x) p_2(x) = 0 \pmod
{p(x)}$ unless either $p_1(x)=0$ or $p_2(x)=0$. The field property 
allows us to prove that the hash function is pairwise independent (see Lemma~\ref{uniformlemma}),
but it is not 3-wise independent because of Proposition~\ref{threewiseprop}.
There is also a direct argument:
\[x(h(x_1,x_1,x_2)-h(x_1,x_1,x_1))+h(x_1,x_1,x_1)=h(x_1,x_2,x_1).\]}
In the sense of Proposition~\ref{threewiseprop}, it is an optimal recursive hashing function.

\begin{lemma}\label{uniformlemma}
 \textsc{General} is pairwise independent.
\end{lemma}
\begin{proof}
If $p(x)$ is irreducible, then $q(x)\in \textrm{GF}(2)[x]/p(x)$ has an inverse, noted $q^{-1}(x)$ since
 $\textrm{GF}(2)[x]/p(x)$ is a field. 
  Interpret hash values as polynomials in $\textrm{GF}(2)[x]/p(x)$.


Firstly, we prove that  \textsc{General} is uniform. In fact,
we show a stronger result: $P(q_1(x) h_1(a_1) + q_2(x) h_1(a_2)+\ldots+q_n(x)h_1(a_n)=y)= 1/2^L$
for any set of polynomials $q_i$ with at least one of them different from zero.
 The result is shown by induction on the number of non-zero polynomials: it is clearly true
where there is a single non-zero polynomial. Suppose it is true for up to $k-1$~non-zero
polynomials and consider a case where we have $k$~non-zero polynomials. 
Assume without loss of generality that
$q_1(x)\neq 0$, we have
$P(q_1(x) h_1(a_1) + q_2(x) h_1(a_2)+\ldots+q_n(x)h_1(a_n)=y)= 
P( h_1(a_1)  = q_1^{-1}(x)(y - q_2(x) h_1(a_2)-\ldots-q_n(x)h_1(a_n)))
=\sum_{y'} P( h_1(a_1)  = q_1^{-1}(x)(y - y'))P(q_2(x) h_1(a_2)+\ldots+q_n(x)h_1(a_n)=y')
= \sum_{y'} \frac{1}{2^L}\frac{1}{2^L}=\frac{1}{2^L}$ by the induction argument. Hence the
uniformity result is shown.

Consider two distinct sequences $a_1,a_2,\ldots,a_n$ and $a'_1,a'_2,\ldots,a'_n$.
 We have that
$P(h(a_1,a_2,\ldots,a_n)=y \land h(a'_1,a'_2,\ldots,a'_n)=y')=P(h(a_1,a_2,\ldots,a_n)=y|h(a'_1,a'_2,\ldots,a'_n)=y')P(h(a'_1,a'_2,\ldots,a'_n)=y')$.
Hence, to prove pairwise independence, it suffices to show that  
 $P(h(a_1,a_2,\ldots,a_n)=y|h(a'_1,a'_2,\ldots,a'_n)=y')=1/2^L$. 

Suppose that
$a_i= a'_j$ for some $i,j$; if not, the result follows since by the (full) independence
of the hashing function $h_1$, the values $h(a_1,a_2,\ldots,a_n)$
 and $h(a'_1,a'_2,\ldots,a'_n)$ are independent.
Write $q(x)= -(\sum_{k | a_k= a_i}  x^k ) (\sum_{k | a'_k= a'_j}  x^k )^{-1}$,
then 
$h(a_1,a_2,\ldots,a_n)+q(x)h(a'_1,a'_2,\ldots,a'_n)$
is independent from $a_i= a'_j$ (and $h_1(a_i)=h(a'_j)$).

In $h(a_1,a_2,\ldots,a_n)+q(x)h(a'_1,a'_2,\ldots,a'_n)$,
only hashed values 
 different from $h_1(a_i)$ remain: label
them $h_1(b_1),\ldots, h_1(b_m)$. The result of the substitution
can be written $h(a_1,a_2,\ldots,a_n)+q(x)h(a'_1,a'_2,\ldots,a'_n) = \sum_k q_k(x) h_1(b_k)  $ where $q_k(x)$ 
are polynomials in $\textrm{GF}(2)[x]/p(x)$. All $q_k(x)$ are zero if and only if $h(a_1,a_2,\ldots,a_n)+q(x)h(a'_1,a'_2,\ldots,a'_n)=0$
for all values of $h_1(a_1),\ldots, h_1(a_n)$ and $h_1(a'_1),\ldots, h_1(a'_n)$
(but notice that the value $h_1(a_i)=h(a'_j)$ is irrelevant);
in particular, it must be true when $h_1(a_i)=1$ and $h_1(a'_i)=1$ for all $i$, hence
$(1+x+\ldots+x^n)+q(x)(1+x+\ldots+x^n)=0\Rightarrow q(x)=-1$. Thus,
all $q_k(x)$ are zero if and only if $h(a_1,a_2,\ldots,a_n)=h(a'_1,a'_2,\ldots,a'_n)$
for all values of $h_1(a_1),\ldots, h_1(a_n)$ and $h_1(a'_1),\ldots, h_1(a'_n)$ which only
happens if the sequences $a$ and $a'$ are identical.
Hence, not all $q_k(x)$ are zero.

 The condition $h(a'_1,a'_2,\ldots,a'_n)=y'$ can be rewritten as 
 $h_1(a'_j) = (\sum_{k | a'_k= a'_j}  x^k )^{-1} (y'- \sum_{k | a'_k\neq a'_j}  x^k h_1(a'_k))$, 
and this last condition is clearly independent from  
$h(a_1,a_2,\ldots,a_n)+q(x)h(a'_1,a'_2,\ldots,a'_n)=y+q(x)y'$ where $h_1(a'_j)=h_1(a_i)$
does not appear.
We have
\begin{eqnarray*}
\lefteqn{P(h(a_1,a_2,\ldots,a_n)=y|h(a'_1,a'_2,\ldots,a'_n)=y')}&&\\
& = &P(h(a_1,a_2,\ldots,a_n)+q(x)h(a'_1,a'_2,\ldots,a'_n)=y+q(x)y')\\
& = &P(\sum_k q_k(x) h_1(b_k) = y + q(x) y')
\end{eqnarray*}
and
by the earlier uniformity result, this last probability is equal to $1/2^L$.
This 
concludes the proof.~\end{proof}

Of the four
recursive hashing functions investigated by Cohen, these two were
superior both in terms of speed and uniformity, 
though \textsc{Cyclic} had a small edge over  \textsc{General}.
For $n$ large, the benefits of these recursive hash functions
compared to the $n$-wise independent hash function presented earlier 
can be substantial: $n$ table look-ups\footnote
{Recall we assume that $\Sigma$ is not known in advance.  Otherwise
for many applications, each table lookup could be merely an array 
access.} is much
more expensive than a single look-up followed by binary shifts.
\komment{
\owen{believes that a couple of array accesses and xors is probably nearly
competitive, since he does not see the need for mini hash tables.}
}

A variation of the Karp-Rabin hash method is ``Hashing by
Power-of-2 Integer Division''~\cite{cohenhash}, where
$h(x_1,\ldots,x_n) = \sum_i x_i B^{i-1} \bmod{2^L}$.  
\notconference{Parameter $B$ needs to
be chosen carefully, so that the sequence $B^k \bmod {2^L}$ for $k=1, 2,\ldots$
does not repeat quickly.}
In particular, the \texttt{hashcode} method
for the Java 1.5 String class uses this approach, with $L=32$ and 
$B=31$~\cite{j15doc:String}. Note that $B$ is much smaller than the range
of values that the 16-bit Unicode characters can assume. 
A widely used textbook~\cite[p. 157]{weis:dsaaj} recommends
a similar Integer Division hash function for strings
with $B=37$  (e.g., $\sum_i x_i 37^{i-1} \bmod R$) and 
hints that R should be prime. 
Since such Integer Division hash functions are  recursive,
quickly computed, and widely used, it is interesting to seek a 
randomized version of them. Assume that $h_1$ is random hash function 
over symbols uniform in $[0,2^L)$, then
define $h(x_1,\ldots,x_n)=h_1(x_1)+B h_1(x_2)+ B^2 h_1(x_3)+\ldots + B^{n-1} h_1(x_n)\pmod{2^L}$
for some fixed integer $B$. We choose $B=37$ (calling the resulting
randomized hash ``ID37''). \notconference{Observe
\owen{I wrote a perl program.  hope it was correct, because I
think Cohen was a bit wrong.  Anyway, I always got a cycle
length of $2^{L-2}$ for $L \in [3,10]$} that it has
a long cycle when $R = 2^L$.}
We do not require $h_1(x_i)\in [0,B)$.  

Observe that ID37 is recursive over $h_1$.
Moreover, by letting $h_1$ map symbols
over a wide range, we intuitively can reduce the undesirable dependence between
$n$-grams sharing a common suffix.  However, in doing so we destroy a
more fundamental property: uniformity.

 \daniel{ So, maybe this says you would have been
 better off choosing $B=38$, say. Maybe. }
The problem with ID37 is shared by all such randomized Integer-Division 
hash functions that map $n$-grams to $[0,2^L)$.  However, they are 
more severe for certain combinations of $B$ and $n$:

\owen{wording fix needed here, to avoid the $n=1$ case!}

\begin{proposition}
Randomized Integer-Division ($2^L$) hashing with $B$ odd is not uniform for $n$-grams, if $n$ is even.
Otherwise, it is uniform, but not pairwise independent.
\end{proposition}

\begin{proof}

\justconference{Omitted; see \cite{viewsizetechreport}.}
\notconference{
We see that $P(h(a^{2k}) = 0) > 2^{-L}$ since
$h(\texttt{a}^{2k}) = h_1(\texttt{a}) ( B^0(1+B) + B^2(1+B) + \ldots + B^{2k-2}(1+B)) \bmod 2^L$
and since $(1+B)$ is even, we have
$P(h(\texttt{a}^{2k}) = 0) \geq P(h_1(x_1)=2^{L-1} \lor h_1(x_1)=0)= 1/2^{L-1}$.  

For the rest of the result, we begin with $n=2$ and $B$ even.
\begin{eqnarray*} 
P(h(\texttt{ab})=y) & = & P(B h_1(\texttt{a})+h_1(\texttt{b})=y\bmod 2^L)\\
&=& \sum_z P(h_1(\texttt{b})=y-Bz \bmod 2^L) P(h_1(\texttt{a})=z)\\
&=& \sum_z P(h_1(\texttt{b})=y-Bz \bmod 2^L)/2^L=1/2^L, 
\end{eqnarray*}
whereas 
$P(h(\texttt{aa})=y)=P((B+1)h_1(\texttt{a})=y \bmod 2^L)=1/2^L$
since $(B+1) x =y \bmod 2^L$ has a unique solution $x$ when $B$ is even. 
Therefore $h$ is uniform. This argument can be extended for any value of $n$ 
and for $n$ odd, $B$ even.
\daniel{Proof by hand waving, but I believe in it.}

To show it is not pairwise independent, first suppose that
$B$ is odd.  For any string $\beta$ of length $n-2$, consider
$n$-grams $w_1 = \beta \texttt{aa}$ and $w_2 = \beta \texttt{bb}$
\notconference{.}\justconference{for distinct
$\texttt{a}, \texttt{b} \in \Sigma$.}
Then 
{\small
\begin{eqnarray*}
P(h(w_1) = h(w_2)) &=& P(B^2 h(\beta) + B  h_1(\texttt{a})+h_1(\texttt{a})
= B^2 h(\beta) + B h_1(\texttt{a})+h_1(\texttt{a}) \bmod 2^L) \\
&=& P( (1+B) (h_1(\texttt{a})-h_1(\texttt{b})) \bmod 2^L = 0) \\
&\geq&  
P(h_1(\texttt{a})-h_1(\texttt{b}) = 0)+P(h_1(\texttt{a})-h_1(\texttt{b})
= 2^{L-1})\\
&=&2/4^L.
\end{eqnarray*}
}
Second, if $B$ is even, a similar argument shows
$P(h(w_3) = h(w_4)) \geq 2/4^L$, where
$w_3 = \beta\texttt{aa}$ and $w_4 = \beta\texttt{ba}$.
}
\end{proof}

These results also hold for any integer-division hash where the modulo
is by an even number, not necessarily a power of 2.  Frequently,
such hashes compute their result modulo a prime.  However, even if
this gave uniformity, the GT algorithm implicitly
applies a ``MOD $2^L$'' operation because it ignores higher-order bits.  
It is easy
to observe that if $h(x)$ is uniform over $[0,p)$,
with $p$ prime, then $h'(x) = h(x) \bmod 2^L$ cannot be uniform.

Whether the lack of uniformity and pairwise independence  is just a
theoretical defect can be addressed experimentally.

\komment{  not for now.
\owen{ For ngrams, suppose we want to compute $n-1$ grams, $n-2$ grams
down to 1-grams.  In that case, it might be handy to have some
kind of incremental updater.  At each position $k$ in the stream,
we compute first $h(x_k)$, then incrementally update it to $h(x_k,x_{k-1})$,
then $h(x_k,x_{k-1},x_{k-2})$,\ldots,
$h(x_k,x_{k-1},x_{k-2},\ldots)x_{k-n+1})$.  The lengths of
the ``grams'' are increasing.  I presume it is fine to reverse the
ngrams for hashing convenience.  So, if the hashing function is like a
stream checksummer the update is cheap.  The bad news here is that
I don't need a recursive hash to compute one n gram hash from another
(adding one char and dropping another),
I need a semi-recursive (made up term) hash that can incrementally 
add a character.  IF this is efficient (O(1) per update), there is no
better way of computing all $n$, $n-1$, \ldots hashes.\\
It gets a bit more interesting if you want to calculate a few
$n$-gram hashes simultaneously: say $n=5,6,7,8,\ldots, 10$.
Well, you can always store the old 5-gram, 6-gram, 7-gram hash
and update each.  If it's O(1) to update, you cannot beat this...
uninteresting.  However,   if you can store the old 5-gram hash,
update it in to the new 5-gram hash, then update this to the
6-gram, 7-gram etc hashes, that's interesting (if it's faster).
}
\daniel{This is interesting, though I'm not 100\% sure I got all of it.
A recursive hash is automatically
semi-recursive. Do you have an example of a semi-recursive hash which 
is not recursive? Also, in your example,  you always begin at
$x_k$, so you'll need recursivity if you want to move on to 
$n$-grams starting at $x_{k+1}$.}
\owen{Yes, that is what I was trying to say....IF you only want,
let's say,  $n/2$-grams  to $n$-grams.  You'll need recursivity
for $n/2$-grams to move on to the next position.  BUT if you plan
to calculate 1-grams, 2-grams, \ldots $n$-grams at the new position,
you DON'T need recursivity, because you can start afresh at each
new position (with the 1-gram).\\
I don't have a concrete example of a semi-recursive hash that is not
recursive, but maybe some of the checksumming routines would be like
that.  With 1.5s thought, a good candidate might be some checksum used
for crypto purposes.\\
I wasn't sure that a recursive hash would be automatically semi-recursive.
It seemed like it ought to be, though.  Do you have a convincing 
explanation other than something like 
``you can tell it the new character and claim 
that the shifted-out character had value 0?''
}
\daniel{No, I don't have a proof, good point.}
\owen{Since Cohen's functions work so well, I don't think we can
get any mileage out of this approach.  So I think we should drop
it and just suppose that we have a separate set of buffers
for each $n$.  In the event that $\Sigma$ is large (eg word ngrams), 
the storage for the character hashers $h_1$ might be high (when totalled up).
There's no reason that the character hashers could not be shared between
different $n$ computations, though.\\
The exception is $n$-wise hashing, where we run ``--slow''.  Yet clearly if
character hashers $h_1 \ldots h_n$ are shared by the $n$-gram estimator
and the $n+1$-gram estimator, it is trivial to xor $h_{n+1}(new char)$ to
the $n$-gram hash to get the $n+1$-gram hash.
If this is valid and your proof that $n>2$-wise cannot be recursive
holds, this seems to warrant experiments--- but I am concerned about
the timeframe.} 
}

\section{Count Estimation by Probing}

Count estimates, using the algorithms of Gibbons and Tirthapura or of
Bar-Yossef et al., depend heavily on the hash function
used and the buffer memory allocated\notconference{ to the algorithms}~\cite{Gibbons2001,BarYossef2002}. 
This section shows that better accuracy bounds from a single run 
of the algorithm follow
if the hash function is drawn
from a family of $k$-wise independent hash functions ($k > 2$), than
if it is drawn from a family of merely pairwise independent functions.
In turn, this implies that less buffer space can achieve
a desired quality of estimates.

Another method for improving the estimates of these techniques is to run them
multiple times (with a different hash function chosen from its family each time).
Then, take the median of the various estimates.
\notconference{
For estimating
some quantity $\mu$ within a relative error (``precision'') of $\epsilon$, it is enough to have 
a sequence of random variables $X_i$ for $i=1,\ldots,q$ such
that $P(\vert X_i - \mu \vert > \epsilon \mu) < 1/3$ where $\mu = \bar X_i$.
The median of all $X_i$ will lie outside $(\mu-\epsilon \mu, \mu+ \epsilon \mu)$
only if more than half the $X_i$ do. This, in turn, can be made
very unlikely simply by considering many different random variables ($q$ large).
Let $Y_i$ be the random variable taking the value 1 when 
$\vert X_i - \mu \vert > \epsilon \mu$ and zero otherwise, and
furthermore let $Y=\sum_{i=1}^q Y_i$.
We have that $E(Y)\leq q/3$ and so, $3 E(Y) / 2 \leq q/2$
Then a Chernoff bound says that~\cite{jfccs174} 
\begin{eqnarray*}
P(Y > q/2)&\leq &P(Y>\bar Y (1+1/2))\\
&\leq& \left (\frac{e^{1/2}}{(1+1/2)^{1+1/2}} \right)^{\bar Y}
\\
&\leq& \left (\frac{e^{1/2}}{(1+1/2)^{1+1/2}} \right)^{q/3}
\\
&\leq& e^{-\frac{q}{10\times 3}}.
\end{eqnarray*} 
Choosing $q=30 \ln 1/\delta$, we have
$P(Y > q/2) \leq \delta$ proving that we can make the median of
the $X_i$'s within $\epsilon \mu$ of $\mu$, $1-\delta$ of the time for $\delta$ arbitrarily
small. }
On this basis, Bar-Yossef et al.~\cite{BarYossef2002}
report that they can estimate a count with relative precision $\epsilon$ and
reliability\footnote{i.e., the computed count is within $\epsilon \mu$ of the
true answer $\mu$, $1-\delta$ of the time} $1-\delta$,
using $O(\left ( \frac{1}{\epsilon^2}+\log m \right)\log \frac{1}{\delta})$ bits
of memory and $O(\left (\log m + \frac{1}{\epsilon}\right )\log \frac{1}{\delta})$ amortized
time. 
Unfortunately, in practice, repeated probing is not a competitive solution
since it implies rehashing all $n$-grams
$30 \ln 1/\delta$ times, a critical bottleneck.
Moreover, in a streaming context,  the various runs are made in parallel and therefore 
$30 \ln 1/\delta$ different buffers are needed.   Whether this is problematic
depends on the application and the size of each buffer.  For $n$-gram estimation,
in one pass we may be attempting to compute estimates for various values of $n$, each of which would
then use a set of buffers.

\komment{
\owen{Weren't the Bar\_yossef people just storing the hashes?  I think you
may've mentioned this, but it's not mentioned here?}
\daniel{
Yes, you are right, they store the hashes thus saving a bit of space, I think, though
I have not fully groked that part of their paper, I focused on the proof of the accuracy
itself. I think storing the hashes instead of the objects is only a small variant, I'm
not sure it buys them much, it might just help a bit, but my claim is that, compared to
having to rehash repeatedly, their savings are small. In effect, I claim that their
complexity analysis is misleading possibly because they have hidden constant factors.}
}

\daniel{Needs some filling here.}

\begin{figure}
\centering
\notconference{\includegraphics[width=.6\columnwidth,angle=270]{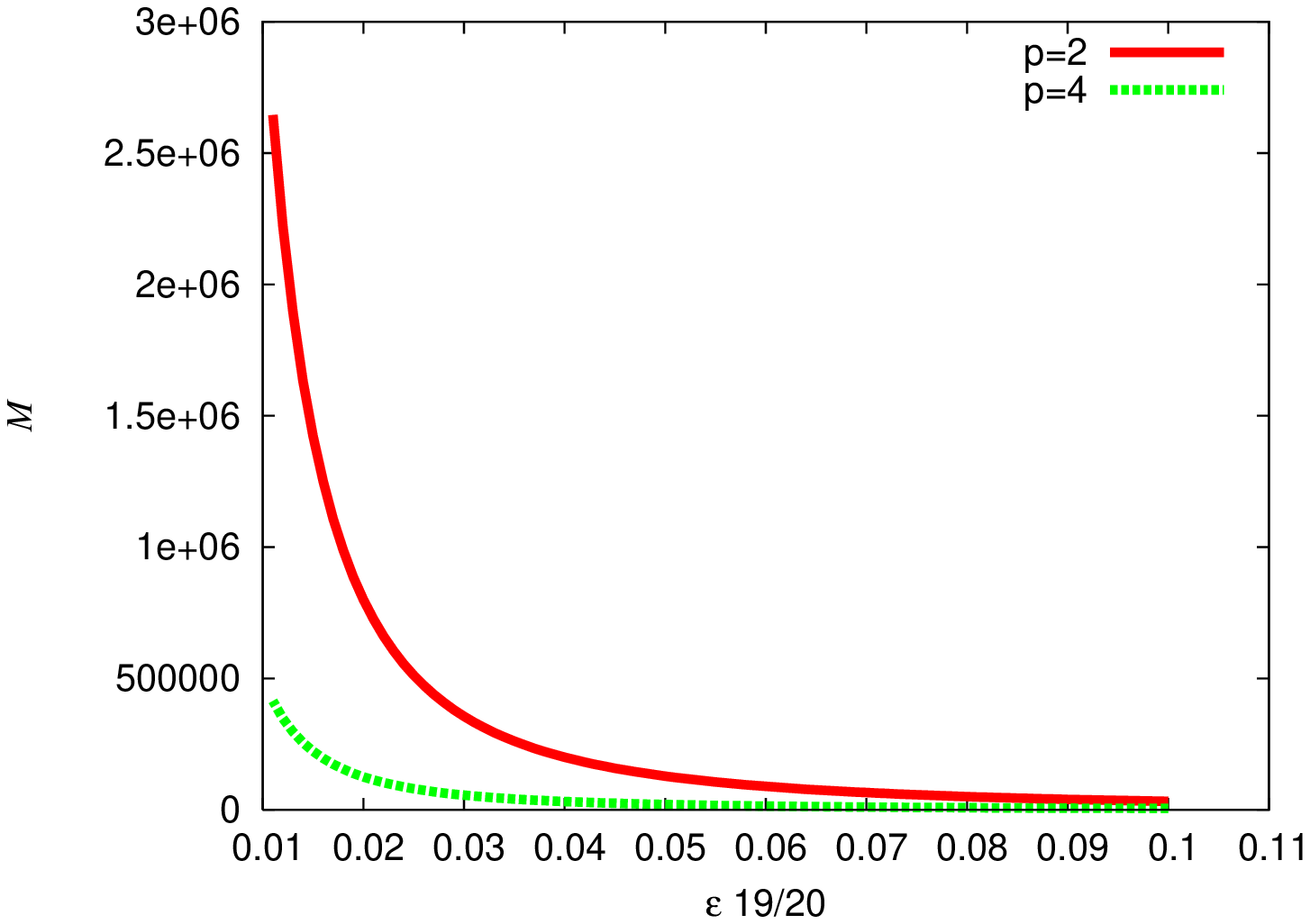}}
\justconference{\includegraphics[width=.4\columnwidth,angle=270]{M-vs-epsilon.eps}} 
 \caption{\label{countpropfig}For reliability $1-\delta = 0.95$, we plot the memory usage $M$ versus
the accuracy $\epsilon$ for pairwise ($p=2$) and 4-wise ($p=4$) independent hash functions
as per the bound of Proposition~\ref{countprop}. \notconference{Added 
independence substantially improves memory  requirements for a fixed estimation accuracy
 according to our
theoretical bounds. }
}
\end{figure}

The next proposition shows that in order to reduce the memory usage drastically,
we can increase the independence of the hash functions.
In particular, we can estimate the count within 10\%, 19 times out of 20 by 
storing respectively 10,500 and 2,500, and 2,000 
$n$-grams\notconference{\footnote {in some dictionary, 
for instance in a hash table}}
depending on whether
we have pairwise-independent, $4$-wise independent 
or $8$-wise independent hash values.
Hence, there is no need to hash the 
$n$-grams more than once if we can assume that hash values are $\approx 4$-wise
independent in practice (see Fig.~\ref{countpropfig}).

\begin{theorem}\label{missinglink}
Let $X_1,\ldots,X_m$  be a sequence of $p$-wise independent random variables that satisfy 
$\vert X_i-E(X_i) \vert \leq 1$. Let $X= \sum_i X_i$, then for $C= \max(p,\sigma^2(X))$, we have
\[P(| X- \bar X | \geq T) \leq \left ( \frac{p C}{e^{2/3} T^2} \right ) ^{p/2}.\] In particular,
when $p=2$, we have 
\[P(| X- \bar X | \geq T) \leq  \frac{2 C}{e^{2/3} T^2 }.\]
\end{theorem}
\begin{proof}
 See Schmidt, Siegel, and Srinivasan, Theorem~2.4, Equation~III~\cite{schmidt1993chb}.
\end{proof}

The following proposition is stated for $n$-grams, but applies generally
to arbitrary items.

\begin{proposition}\label{countprop}Hashing each $n$-gram
only once, we can estimate the number of distinct $n$-grams 
within relative precision $\epsilon$, 
with a  $p$-wise independent hash for $p\geq 2$ by storing $M$ distinct $n$-grams ($M\geq 8 p$)
and  with reliability $1-\delta$ where $\delta$ is given by 
\[\delta= \frac{p^{p/2}}{ e^{p/3}  M^{p/2}} \left (  2^{p/2}
+\frac{ 8^{p/2}}{ \epsilon^p  (2^{p/2}-1)} \right ).\]
More generally, we have
\begin{eqnarray*}\delta
  &\leq & \frac{p^{\frac{p}{2}}}{ e^{\frac{p}{3}}  M^{\frac{p}{2}}  } \left ( \frac{\alpha^{\frac{p}{2}}}{(1-\alpha)^p}+ \frac{4^{\frac{p}{2}}}{\alpha^{\frac{p}{2}} \epsilon^p (2^{\frac{p}{2}}-1)}\right ) .
\end{eqnarray*}
for $4p/M\leq \alpha<1$ and any $p,M$.
\end{proposition}
\begin{proof}

We generalize a proof by Bar-Yossef et al.~\cite{BarYossef2002}
Let $X_t$ be the number of distinct elements having only zeros in the
first $t$ bits of their hash value. Then $X_0$ is the number of
distinct elements ($X_0=m$). For $j=1,\ldots,m$, let $X_{t,j}$ be the
binary random variable with value 1 if the $j^{\textrm{th}}$ distinct
element has only zeros in the first $t$ bits of its hash value, and
zero otherwise. We have that $X_t = \sum_{j=1}^{m} X_{t,j}$ and so
$E(X_t)= \sum_{j=1}^{m} E(X_{t,j})$.  Since the hash function is uniform,
then $P(X_{t,j}=1)=1/2^t$ and so $E(X_{t,j})=1/2^t$, and hence,
$E(X_t)=m/2^t \Rightarrow 2^t E(X_t)=m$. Therefore $X_t$ can be used
to determine $m$.

Using pairwise independence, we can can show that
$\sigma^2(X_t) \leq \frac{m}{2^t}$. 
\justconference{(For details, see 
\cite{viewsizetechreport}.)}
\notconference{
We have that $\sigma^2(X_t) \leq \frac{m}{2^t}$ because
\begin{eqnarray*}
E(\vert X_t - \frac{m}{2^t}\vert^2) & = &
E( (X_t - \frac{m}{2^t})^2 )\\
& =  & 
E( (\sum_{j=1}^{m}  ( X_{t,j} - \frac{1}{2^t} ))^2  ) = \sigma^2(\sum_{j=1}^{m} X_{t,j})\\
& =  & 
\sum_{j=1}^{m} \sigma^2( X_{t,j}) \mbox{\ (by $2$-wise independence)}\\
& =  & 
\sum_{j=1}^{m} \frac{1}{2^t} (1-  \frac{1}{2^t}) ^2+ (1-\frac{1}{2^t})(0 -  \frac{1}{2^t})^2 \\
& \leq  & 
\sum_{j=1}^{m} \frac{1}{2^t} =\frac{m}{2^t}.
\end{eqnarray*}
}
By
 Theorem~\ref{missinglink}, 
 we have \[P(\vert X_t - m/2^t \vert \geq \epsilon m /2^t) \leq \frac{ p^{p/2} 2^{tp/2}}{m^{p/2} \epsilon^p e^{p/3}}, \mbox{~as long as 
$\sigma^2(X_t)\geq p$.}\]

Let $M$ be the available memory and suppose that the hash function has $L$ bits
such that we know, a priori, that $P(X_L > M)\approx 0$. This is necessary since
we do not wish to increase $L$ dynamically. 
It is reasonable since for $L$ and $m$ fixed, $P(X_L > M)$ goes down as $1/M^p$:
\begin{eqnarray*}P(X_L > M)& =&P(X_L-m/2^L > M-m/2^L)\\
 &\leq & P(\vert X_L-m/2^L \vert > M-m/2^L) \\
&\leq & \left (\frac{p \max(m/2^L,p) }{ e^{2/3} (M-m/2^L)^2} \right )^{p/2}
\end{eqnarray*}
for $M-m/2^L >0$ where we used Theorem~\ref{missinglink}.
For example, if $p=2$, $M=256$, $L=19$, $P(X_L>M)\leq 4.4 \times 10^{-8}$ for $m=100,000,000$.

The algorithm
returns the value $2^{t'} X_{t'}$ where $t'$ is such that $X_{t'} \leq M$
and $X_{t'-1} > M$.  
\notconference{(For a given hash function, note that
$X_i$ is monotone in $i$ and thus $t'$ is uniquely determined
by the input and the hash function.)  
The upshot is that}\justconference{Note that}
$t'$ is itself a random quantity that depends
deterministically on the hash function and the input (the same
factors that determine $X_t$.)
\owen{That needs to be said better.  I think it needs to be said, because
I was confused for a while about why this $t'$ thingie seemed to be
an unknown quantity between 0 and L.  }

We can bound the error of this estimate as follows.
\justconference{
(See \cite{viewsizetechreport} for intermediate steps in the derivation.)
}

\begin{eqnarray*}
\lefteqn{P( \vert 2^{t'} X_{t'} - m\vert \geq \epsilon m)}\\
\notconference{& = &  \sum_{t=0,\ldots , L} 
P( \vert X_t - \frac{m}{2^t}\vert \geq \epsilon \frac{m}{2^t}) P(t'=t)\\}
& = &\sum_{t=0,\ldots , L} 
P( \vert X_t - \frac{m}{2^t}\vert \geq \epsilon \frac{m}{2^t}) P(X_{t-1}>M, X_{t}\leq M)
\notconference{.
\end{eqnarray*} 
Splitting the summation in two parts, we get

\begin{eqnarray*}
\lefteqn{P( \vert 2^{t'} X_{t'} - m\vert \geq \epsilon m)}
 \\
}
& \leq & \sum_{t=0}^{\bar t-1} 
P( \vert X_t - \frac{m}{2^t}\vert \geq \epsilon \frac{m}{2^t})
+ \sum_{t=\bar t ,\ldots , L} 
P(X_{t-1}> M, X_{t}\leq M)\\
& = &  P(X_{\bar t-1}>M)
+ \sum_{t=0}^{\bar t-1} 
P( \vert X_t - \frac{m}{2^t}\vert \geq \epsilon \frac{m}{2^t})
\\
\notconference{& \leq &  P(X_{\bar t-1}-m/2^{\bar t-1}>M-m/2^{\bar t-1}) 
 + \sum_{t=0}^{\bar t-1} 
\frac{ p^{p/2} 2^{tp/2}}{m^{p/2} \epsilon^p e^{p/3}}\\}
& \leq &  P(\vert X_{\bar t-1}-m/2^{\bar t-1} \vert >M-m/2^{\bar t-1}) 
+ \sum_{t=0}^{\bar t-1} 
\frac{ p^{p/2} 2^{tp/2}}{m^{p/2} \epsilon^p e^{p/3}}\\
 \\
\notconference{ 
& \leq & \left ( \frac{pm}{2^{\bar t-1} e^{2/3} (M-m/2^{\bar t-1})^2} \right )^{p/2}
+  \sum_{t=0}^{\bar t-1} 
\frac{ p^{p/2} 2^{tp/2}}{m^{p/2} \epsilon^p e^{p/3}}\\
\\}
& \leq & 
 \frac{p^{p/2} m^{p/2}}{2^{(\bar t-1)p/2} e^{p/3} (M-m/2^{\bar t-1})^{p}} 
+
\frac{p^{p/2}( 2^{ {\bar t} p/2} - 1)}
{m^{p/2} \epsilon^p e^{p/3} (2^{p/2}-1)
}
\end{eqnarray*} 
where we assumed that $p\leq m/2^{\bar t-1}.$

Choose $\bar t \in \{0,\ldots, L\}$ such that $\alpha M/4 < \frac{m}{2^{\bar t}} \leq \alpha M/2$
for some $\alpha <1$ satisfying $\alpha M/4 \geq p$
then 
\[\frac{p^{p/2} m^{p/2}}{2^{(\bar t-1)p/2} e^{p/3} (M-m/2^{\bar t-1})^{p}} \leq 
\frac{p^{p/2} M^{p/2} \alpha^{p/2}}{ e^{p/3} (1-\alpha)^p M^p}
\]
whereas 
\[\frac{p^{p/2}( 2^{ {\bar t} p/2} - 1)}
{m^{p/2} \epsilon^p e^{p/3} (2^{p/2}-1)
} \leq 
\frac{p^{p/2} 4^{p/2}}{ \epsilon^p \alpha^{p/2} M^{p/2} e^{p/3} (2^{p/2}-1)}.\]
Hence, we have
\begin{eqnarray*}P( \vert 2^{t'} X_{t'} - m\vert \geq \epsilon m)
  &\leq & \frac{p^{\frac{p}{2}}}{ e^{\frac{p}{3}}  M^{\frac{p}{2}}  }  ( \frac{\alpha^{\frac{p}{2}}}{(1-\alpha)^p}+ \frac{4^{\frac{p}{2}}}{\alpha^{\frac{p}{2}} \epsilon^p (2^{\frac{p}{2}}-1)} ) .
\end{eqnarray*}
Setting $\alpha=1/2$,
 we have 
\begin{eqnarray*}P( \vert 2^{t'} X_{t'} - m\vert \geq \epsilon m)
  &\leq & 
\frac{p^{p/2}  2^{p/2}}{ e^{p/3}  M^{p/2}}
+\frac{p^{p/2} 8^{p/2}}{ \epsilon^p  M^{p/2} e^{p/3} (2^{p/2}-1)}.
\end{eqnarray*}
For different values of $p$ and $\epsilon$, other values of $\alpha$ can give
tighter bounds: the best value of $\alpha$ can be estimated numerically. The proof
is completed.
~\end{proof}

\notconference{ It may seem that the result of Proposition~\ref{countprop} is independent
of the size of the data set and of the number of distinct symbols ($m$).
Indeed, it provides a fixed accuracy $\epsilon$ for a given  memory budget ($M$) irrespective
of the data source. However, as the proof
shows, the number of bits in the hash values need to grow with $\log m$.
This requires
additional computational costs.
\daniel{...I want to be honest with the reader and admit that we,
and Gibbons, cheat a bit by assuming that $L$ is big enough. How do we know?
Well, the cheat is not significant because of the logarithm involved... nevertheless...}
}

Again the following corollary applies not only to $n$-grams but also to arbitrary items.
It follows from setting $M=576/\epsilon^2$, $p=2$, and $\alpha=1-\epsilon$.
\justconference{(See \cite{viewsizetechreport} for the derivation.)}

\begin{corollary}\label{antibar}
With $M=576/\epsilon^2$ and $p=2$, we can estimate the count of distinct $n$-grams  with 
reliability ($1-\delta$) of 99\% for any $\epsilon>0$.
\end{corollary}
\notconference{
\begin{proof}
 From Proposition~\ref{countprop}, consider
\begin{eqnarray*}\delta
  &\leq & \frac{p^{\frac{p}{2}}}{ e^{\frac{p}{3}}  M^{\frac{p}{2}}  } \left ( \frac{\alpha^{\frac{p}{2}}}{(1-\alpha)^p}+ \frac{4^{\frac{p}{2}}}{\alpha^{\frac{p}{2}} \epsilon^p (2^{\frac{p}{2}}-1)}\right ) 
\end{eqnarray*}
for $\alpha = 1-\epsilon$, $M=576/\epsilon^2$, and $p=2$, then 
\begin{eqnarray*}\delta
  &\leq & \frac{2 \epsilon^2}{ e^{\frac{2}{3}}  576 } \left ( \frac{1-\epsilon}{\epsilon^2}+ \frac{4}{(1-\epsilon ) \epsilon^2 }\right )\\
  &\leq &\frac{2 }{ e^{\frac{2}{3}}  576 } \left ( 1-\epsilon + \frac{4}{(1-\epsilon ) }\right ).
\end{eqnarray*}
Taking the limit as $\epsilon$ tends to 0, on the basis that the probability of a miss 
($P(\vert X_t - m/2^t \vert \geq \epsilon m /2^t) $) can only grow as $\epsilon$ diminishes,
we have that $\delta$ is bounded by  ${10 }/ ({ e^{\frac{2}{3}}  576 })\approx 0.008$.
\end{proof}
}

This significantly improves the bound of 1/6 for $\delta$ given in Bar-Yossef 
et al.~\cite{BarYossef2002}, for the same value of $M$.  

\notconference{
Because $p$-wise independence implies
$p-1$-wise independence, memory usage, accuracy and reliability can only improve as $p$
increases.
For $p$ large ($p\gg 4$), the result of Proposition~\ref{countprop} is no longer useful
because of the $p^p$ factor. Other inequalities on $P(\vert X - \mu \vert )$
for $p$-wise independent $X$'s should be used. 
}

\subsection{Entropy and Iceberg Estimation}
\label{iceberg-entropy}

With only minor modifications, the techniques used for counting can be
modified to provide estimates of various other 
statistics. We consider
entropy~\cite{Guha2006,Chakrabarti2006,Batu2002} 
and iceberg counts~\cite{iceberg98}: the
GT algorithm is modified to track not only the existence
of items hashing to zero, but also the required properties (occurrence
counts, for instance) of these items.  From this, the statistic  can
be estimated for the entire data stream.

Let $I$ be the entire set of $n$-grams
($N=\textrm{card}(I)$) and $I'$ be the set of \textbf{probed}
$n$-grams with $m'=\textrm{card}(I')$ ($m' \in [M/2 , M)$). For $i\in
I'$, we know the exact number of occurrences $f_i$ of $i$ and so the
probability that any given $n$-gram in $I$ is $i$, $P(i)$, is given by
$P(i)=f_i/N$.  Hence, we can compute $\sum_{i\in I'} P(i) \log P(i)$
exactly. A practical one-pass estimate of the Shannon entropy of the
$n$-grams ($\sum_{i\in I} P(i) \log P(i)$) is $m/m' \sum_{i\in I'}
P(i) \log P(i)$ since we have a tight estimate for $m$. Unfortunately,
we do not have any theoretical results on this estimator.
 If we were to
allow two passes over the data, an unassuming algorithm
can estimate the entropy with theoretical bounds~\cite{Guha2006}.
It first samples and then probes.

Various ``iceberg count'' properties can be handled similarly.  In these 
problems, one is given a predicate on the number of occurrences $f_i$ of an item 
$i$. We seek to estimate the number
of distinct items satisfying the predicate.
For example, if the predicate is ``$f_i > c$ for some fixed $c$'', then
we wish to estimate $\textrm{card}(\{i\in I | f_i > c\})$.
For instance, the input text \texttt{aabaabb} contains the
2-grams \texttt{aa} (with count 2), \texttt{ab} (with count 2),
\texttt{ba} (with count 1) and \texttt{bb} (with count 1).
If the predicate $Q(i)$ is ``item $i$ occurs exactly twice'', then
the answer to the iceberg query is 2 (because distinct items
\texttt{aa} and \texttt{ab} satisfy $Q(i)$.  
Considering the predicate ``$f_i > 0$'', the iceberg-count problem
generalizes the problem of counting distinct $n$-grams.

\komment{

I see the SUFARY package has an ngram tool called sang and it
allows a threshold argument.

 probably a novel idea? talk about iceberg cubes
~\cite{iceberg98} they cite \cite{Olken1993} as a good review for ``Random sampling from databases''

they seem to mix probabilistic counting with sampling ugh.

Define iceberg count.

give algorithm we use.

\owen{ Is their idea that you expect to miss many of the things that occur
less than 10 times, if you take a 10\% sample?}
\daniel{I don't recall, sorry. But their approach seemed crude and unlikely
to give great results. I don't recall any benchmarking.}

\owen{ I see your code essentially assumes that the collection of
``high-zeros'' elements is a sample that has count properties 
that mirror those of the entire population.  I worry that, with
high probability, this can break down in the following degenerate
situation: there are one or two extremely common values (whp their
hash mapping won't have many zeros) and zillions of rare values
that occur fewer times than the iceberg threshold.  whp these guys
will be accumulated by GT, and at the end you will conclude
that the iceberg count is 0.  I presume that you do entropy the
same kind of way.  I would think that zipfian data would be almost
a worst case and I am amazed that entropy works well at all...}

\daniel{As far as I know, you are right though I think that 
whatever the competition does might/should be even worse. See
also the following argument:}
}

While we do not have any theoretical bound on the accuracy of such
iceberg estimate, we can still model the problem statistically.
If we are picking at random $m'$ elements from a population of size $m$
having $r<m$ elements satisfying a predicate, the number of elements 
satisfying the constraint, $Y$, follows an 
hypergeometric distribution $Y$ with mean
$m' r/m$ and with variance $m' \frac{r(m-r)(m-m')}{m^2 (m-1)}$. Therefore,
by Chebyshev's inequality, we have that 
$P(\vert Y-m' r/m \vert < \epsilon m' r/m ) \leq  
 \frac{(m-r)(m-m')}{\epsilon^2 m'r (m-1)}< \frac{m}{\epsilon^2 m' r}$. 
Hence,  we should allocate roughly $M\approx \frac{20m}{\epsilon^2 r}$ units of
storage to have an accuracy of $\epsilon$, 19 times out of 20.
Choosing $m'=\frac{200 m}{r}$ ensures that the number
of interesting items found ($Y$) is larger than $190$, 19 times out of 20. Hence,
if an upper bound on $m$ and a non-trivial lower bound on $r$ were known, further theoretical
results could be possible.

\owen{Well, in advance we don't know $r$  or $m$ or $r/m$.  We can estimate
$m$ in a first pass, and also examine the buffered items to estimate
$r/m$.  Then, in a second pass I guess you can try this. Again, you may find
0 items are interesting in your pass1 buffer.
It seems like you might easily wind up with a buffer size that is 
close to $m$, which defeats the spirit of these techniques.
}
\komment{
If one or two items are particularly common in the data stream, unless
$M$ is large, we will probably not probe for those frequent items, and thus
entropy and iceberg count estimates are expected to be poor when data
follows the Zipfian distribution (eg, for $n$-grams of English text). }

If $M$ is small and the distribution is biased (e.g. Zipfian), such
as is the case with $n$-grams of English text, we
expect these entropy and iceberg count estimates to be poor.


\subsection{Simultaneous Estimation}
\label{simultaneous-section}

The GT approach can also be applied to estimate the
number of 1-grams, 2-grams, \ldots $n$-grams in one pass over the data.
Initially, one might seek hash families such that if
$i$ is a $\beta+1$-gram and $j$ is its $\beta$-gram suffix,
then $h(i)$ has zero in its first $t$ bits
whenever the hash value $h(j)$ has zero in its first $t$ bits. For example, 
$h(a)=0 \Rightarrow h(b a) =0$\footnote{We abuse the notation
by using $h$ to denote both
the hash function for bigrams and the hash function for unigrams.}.
Unfortunately, such a family of hash functions cannot be pairwise independent
because if $h(\texttt{ba})=0$ then $h(c\texttt{a})=0$ 
for any value of $c$.
However, a straightforward approach can suffice for
simultaneous estimation. Separate buffers ($n$ in total) can be kept for 
each string length ($1$-gram, $2$-gram, \ldots, $n$-gram), 
and for simplicity we can assume each buffer is of 
the same  size, $M$.

The straightforward approach works as follows:
As each of the $N$ symbols in the stream is processed, we hash
the  $n$ new $n$-grams of size 1, 2, \ldots, $n$ respectively.   If each of the $n$ hashes is recursive, then
each can be updated from its previous value in O(1) time.  However, we
can also use semi-recursive hashes with this approach, because the hash value
for a $k$-gram can be updated, in O(1) time, to become the hash value for
the associated $k+1$-gram.  

In particular, the semi-recursive ``$n$-wise'' hashing scheme examined 
before becomes more attractive when simultaneous estimation is attempted.

\komment{
(WHat follows needs fixing: n-dimensional probing is not n-1
dimensional probing! Something is wrong.)  In practice, though, you'd
need to store the tuples, because you would want to reuse these tuples
for simultaneously estimating not only the n-grams but also the
n-1-grams and so on. Random n-dimensional probing ought to imply
random n-1-dimensional probing! Hence Bar-Yossef is not a good idea
with n-grams.

Ah! AH! So, no, you don't want to store only the hash, because I think
you want to reuse the selected tuples, project them onto smaller cubes
for more estimates. (Show here, mathematically and algorithmically,
how it could work.)

This is kind of having a sample of your data cube and working from that. 

Naturally, you could do simultaneous n-gram, n-1-gram, n-2-gram
estimations from the hashes, so, you hash the 2-gram, then from it the
3-gram and so on, but it requires lots of memory. (Owen claim it could
be made to work, maybe? Discuss!)
}

\komment{

\section{Complexity and Implementation}

\owen{too light to be a section.  But where to jam it?  Into experiments since it
was done so that we could experiment?}

ok, when a new n-gram arrives, much of the time is spent on hashing it.
once it is hashed, probing is inexpensive: most of the time, we only check
whether the hash belongs to the probed set, and this can be done in a time
that depends on the number of bits, which itself is either a fixed and small number
(19).

\owen{isn't this just a variant on the Bar-Yossef ``only store the hashes'' approach?
Did Gibbons suggest this?} \daniel{We didn't use the BY ``only store hashes'' technique, we used straight Gibbons,  as far as I can tell, except that we apply it to $n$-grams, we have tighter theoretical bounds, and we have experimental validation.}
\owen{some code (C++ or Python) stores *both* hashes and ngram.  You first 
compare hashes for identity as a speed optimization and then, when then hashes 
match, you actually compare ngrams.  But the optimization seems very BY-ish, 
unless Gibbons suggested it too.}

\komment{
Hashing performance depends on pseudorandom generator. The standard
C random generator used by GCC is relatively slow and state-of-the-art
random generators such as Mersenne Twister~\cite{matsumoto1998mtd}
have better statistical properties and can be at least 2 to 3 times faster.
\owen{speed may have to be better sales}
And we argue that solution quality isn't either.
}

Written in C++ and Python (any Python survived or are we all C++ now?)

C++ code available
}

\section{Experimental Results}
Experiments are used to assess the accuracy of estimates obtained by
several hash functions on some input streams.  As well, the experiments
demonstrate that the techniques can be efficiently implemented.
\justconference{
Space prohibits a discussion of all the experiments, but details 
are available in \cite{viewsizetechreport}
and summarized below.}%
Our code is written in C++ and is available upon request.

\subsection{Test Inputs}

One complication is that the theoretical predictions are based on
worst-case analysis. There may not be a sequence of symbols realizing
these bounds. 
\komment{However, for a family $\mathcal{H}$ of hash
functions, we do not know how to determine such a worst-case input
theoretically  and enumerating all possible inputs is clearly out of
the question.}%
\daniel{ I kill this, replaced by above.
In fact, the worst-case input may not even correspond to a stream
of $n$-grams (where $n-1$ characters from one $n$-gram reappear in the
following $n$-gram).   As well, we want to know how these techniques
perform on ``typical'' inputs.}%
As a result, our experiments used
the $n$-grams from
a collection of 11~texts\footnote{The 11 texts are eduha10 (The Education of Henry Adams), 
utrkj10 (Unbeaten Tracks in Japan),
utopi10 (Utopia), remus10 (Uncle Remus His Songs and His Sayings), 
btwoe10 (Barchester Towers), 00ws110 (Shakespeare's First Folio), 
hcath10 (History of the Catholic Church), rlchn10 (Religions of Ancient China), 
esymn10 (Essay on Man), hioaj10 (Impeachment of Andrew Johnson), and
wflsh10 (The Way of All Flesh).} from  Project Gutenberg.  We also used synthetic data sets
generated according to various generalized Zipfian distributions.
\owen{Found lots of Zipfian distributions some Zipfian data sets called such
on Google Scholar}
Since we are analyzing the performance of several randomized algorithms,
we ran each algorithm 100+ times on each text.
\notconference{
We cannot
run tests on inputs as large as would be appropriate for corpus linguistics
studies: to complete the entire suite of experiments in reasonable time,
we must limit ourselves to texts (for instance, Shakespeare's First Folio)
where one run takes at most a few minutes.  
}

\subsection{Accuracy of Estimates}

\begin{table}[tb]
\caption{
\label{table:theoretical-error-bounds}
Maximum error rates $\epsilon$ 19 times out of 20 for various amounts of memory ($M$) and for
$p$-wise independent hash values according to Proposition~\ref{countprop}
.}
\begin{center}
\begin{tabular}{|c|cccccc|}
\hline\hline
       & 256     & 1024    & 2048    & 65536  & 262144 & 1048576 \\ \hline
$p=2$  & 86.4\%  & 36.8\%  & 24.7\%  & 3.8\%  & 1.8\%  & 0.9\% \\
$p=4$  & 34.9\%  & 16.1\%  & 11.1\%  & 1.8\%  & 0.9\%  & 0.5\% \\
$p=8$  & 30.0\%  & 14.1\%  & 9.7\%  & 1.6\%  & 0.8\%  & 0.4\% \\
\hline \hline
\end{tabular}
\end{center}
\end{table}

We have theoretical bounds relating to the error $\epsilon$ observed
with a given reliability (typically 19/20), when the hash function
is taken from a $p$-wise independent family.
(See Table~\ref{table:theoretical-error-bounds}.)  But how close to this
bound do we come when $n$-grams are drawn from a ``typical'' input
for a computational-linguistics study?  And do hash functions from
highly independent families actually enable more accurate\footnote
{The ``data-agnostic'' estimate from Sect.~\ref{relatedwork} is
hopelessly inaccurate:
it predicts 4.4~million 5-grams for Shakespeare's First Folio, but
the actual number is 13 times smaller.}
estimates?

\begin{figure}[h]
\begin{center}
\subfigure[Count estimate errors over Shakespeare's First Folio (00ws110), 100 runs estimating 10-grams with $M=2048$.]{
\includegraphics[width=0.39\columnwidth,angle=270]{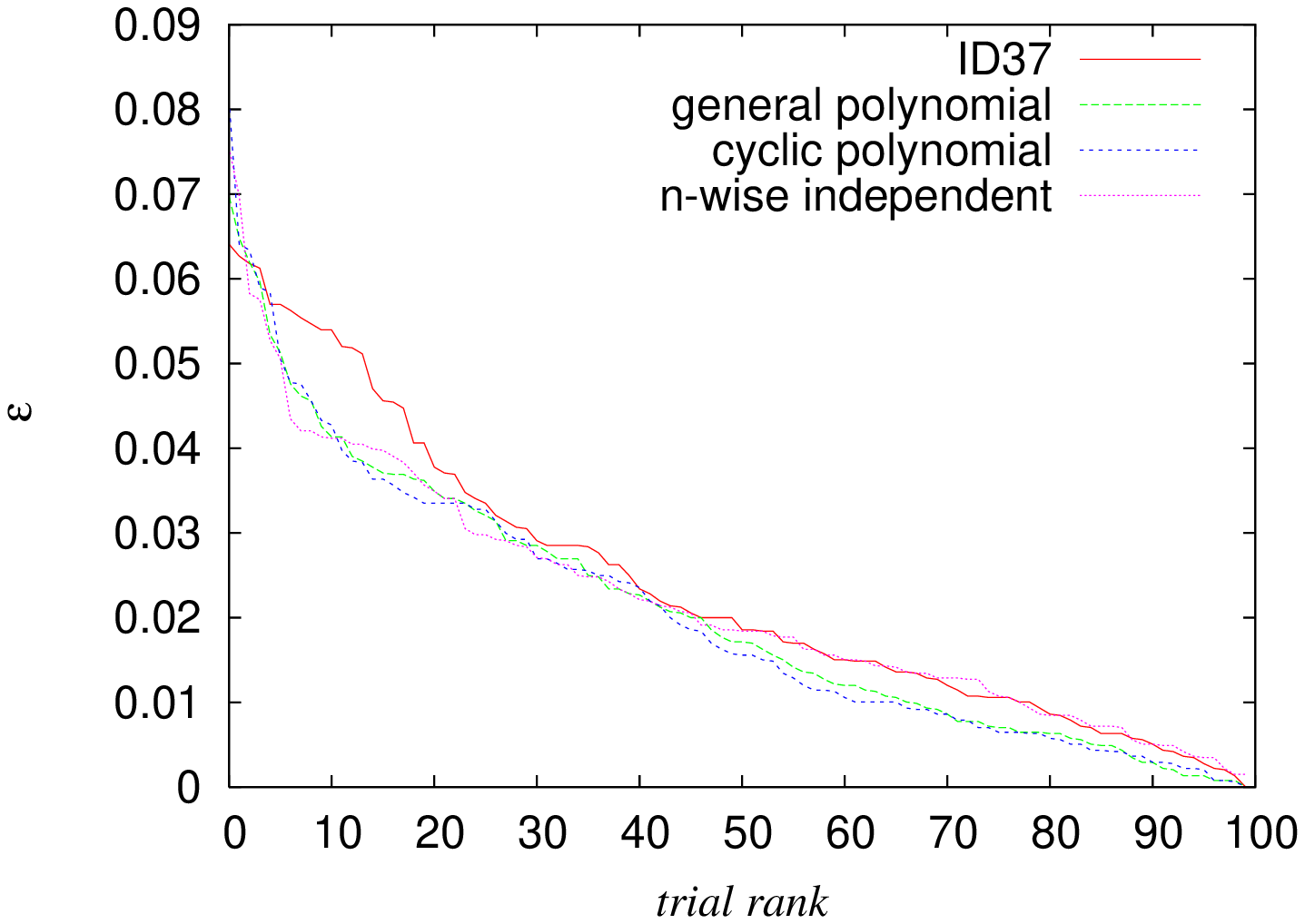}
}
\subfigure[For each $M$ and hash function, worst-case  $95^\textrm{th}$-percentile
error observed on the 11 test inputs]{
\label{shakespeare-twok-ten-errors11}
\includegraphics[width=0.39\columnwidth,angle=270]{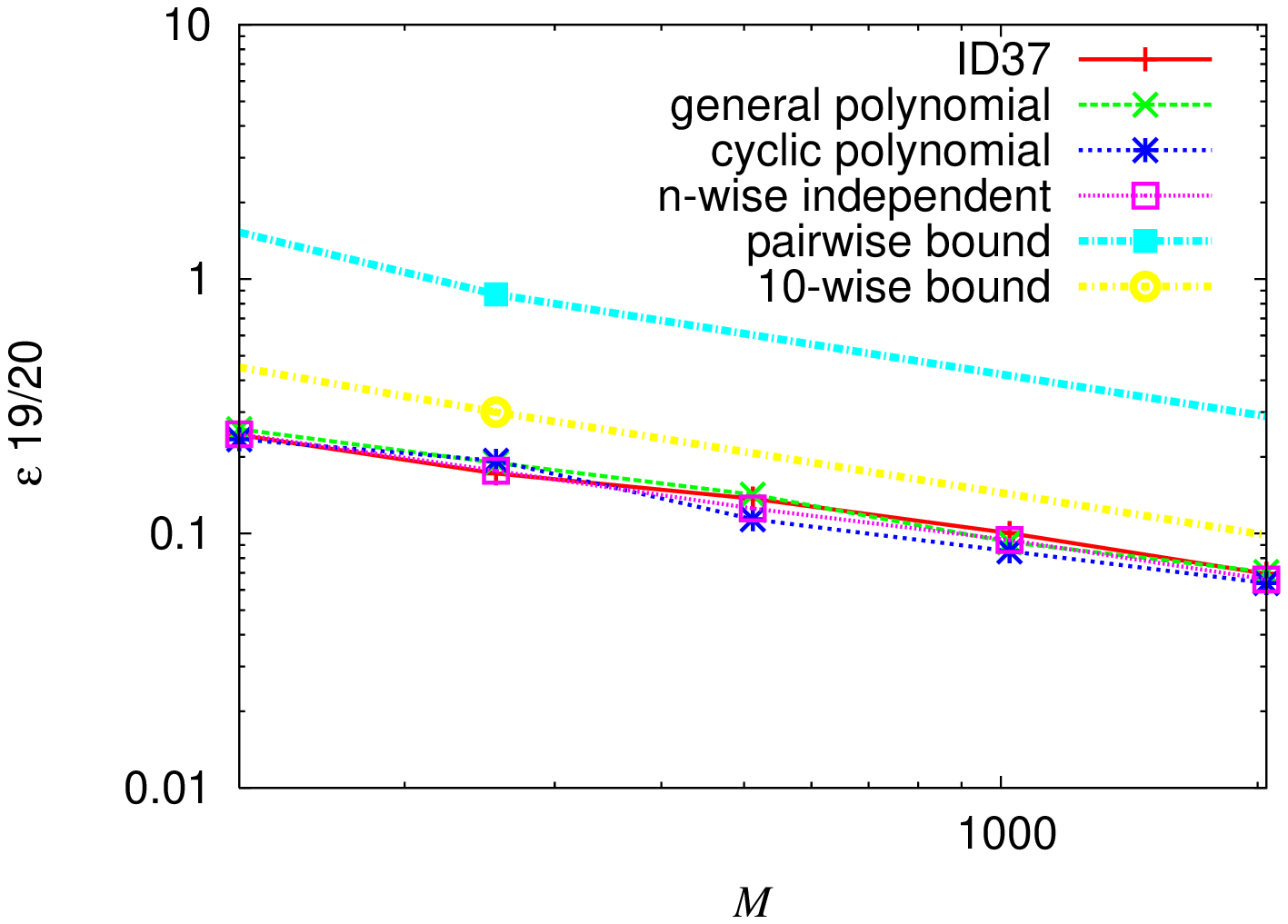}
}
\end{center}
\caption{
\label{shakespeare-twok-ten-errors}Average relative error $\epsilon$ after 100~runs and over four hash functions. }
\end{figure}

Figure~\ref{shakespeare-twok-ten-errors} shows the relative error $\epsilon$
observed from four hash functions (100 estimations with each).  
Estimates have been ranked by decreasing $\epsilon$, and 
we see ID37 had more poorer runs than the others. \komment{ Note also that
most runs are not far worse than the 95th percentile.
Note how, if we selected 9 runs, we'd normally have at least one
really good one.  But I guess it might not necessarily be the
median of the 9!}%
\komment{
arbitrary choice of btowe, not interesting...

{\centering
\includegraphics[width=0.6\columnwidth]{epsilon-M=1024-n=10-btowe.eps}
}
}%
\notconference{
Figure~\ref{remus-twofivesix-ten-errors} shows a test input (remus10) that was
the worst of the 11 for several hash functions, when $M=256$.  ID37 seems
to be doing reasonably well, but 
we see $10$-wise independent hashing lagging.

\begin{figure}
\begin{center}
\includegraphics[width=0.5\columnwidth,angle=270]{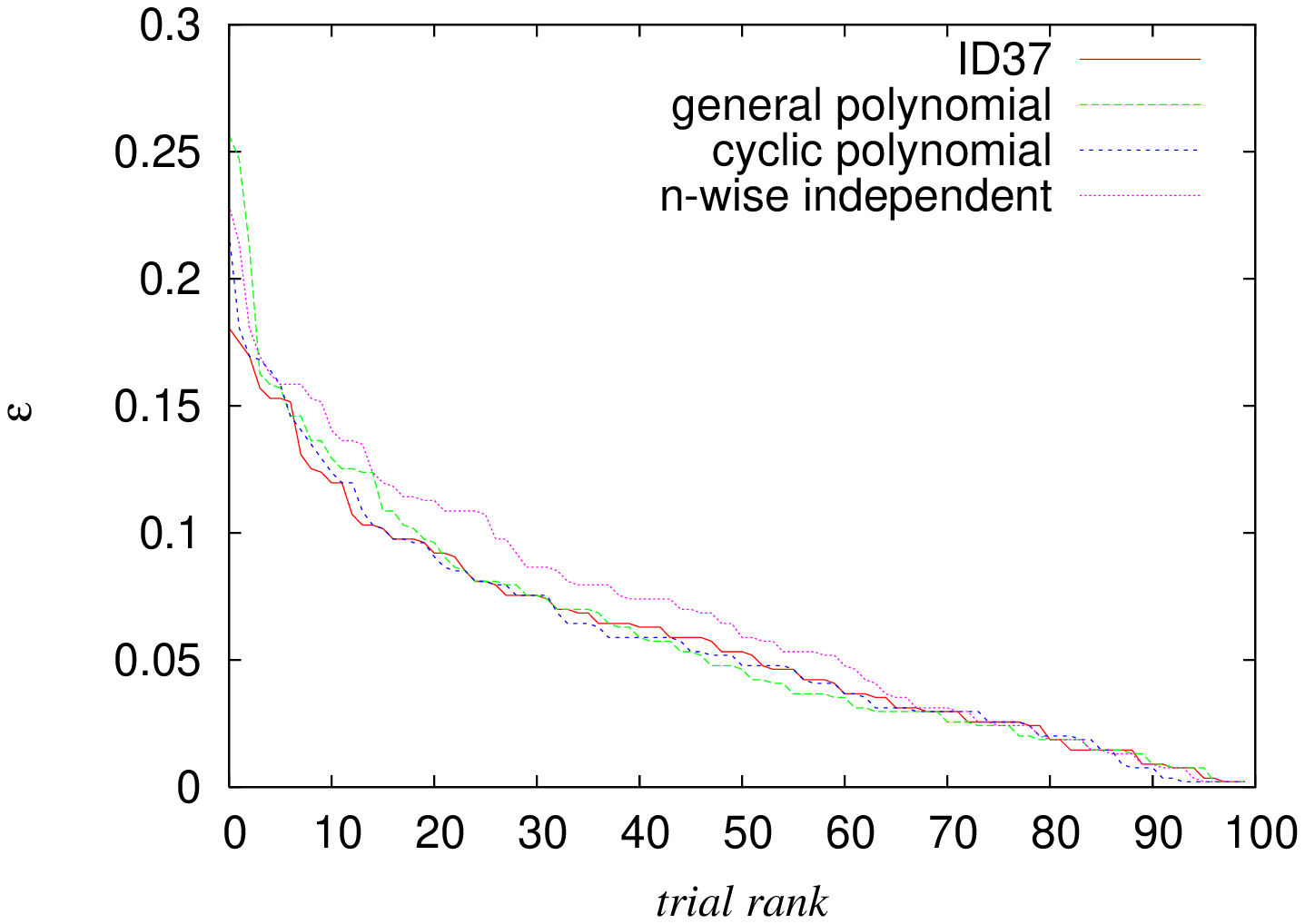}
\end{center}
\caption{
\label{remus-twofivesix-ten-errors}Errors on remus10 (``Uncle Remus His Songs and His Sayings''), from 100 runs estimating 10-grams with $M=256$.}
\end{figure}
}

To study the effect of varying $M$,
we use the 
$5^{\textrm{th}}$-largest error of  100 runs.  This 
$95^\textrm{th}$-percentile error can be related to the theoretical bound 
for $\epsilon$ with 19/20 reliability.
\komment{
\subsection{ $\textbf{95}^\textrm{th}$-Percentile Errors on Selected Gutenberg Texts }
}%
Figure~\ref{shakespeare-twok-ten-errors11}  
plots the largest 
$95^\textrm{th}$-percentile error observed over 11~test inputs.
\komment{
from Project Gutenberg.  Every hash method was run at least 100
times for each selected value of $M$.    Taking the maximum
of the 11, rather than selecting the $95^\textrm{th}$-percentile
of 1100 runs, is inspired by a desire to have a worst-case experimental
analysis.  \komment{While we do not know the theoretical worst-case input, if}
If some of the 11 inputs are harder than others, an effect will be shown.

Even considering the logarithmic $y$-axis display,}%
It is apparent that there is
no significant accuracy difference between the hash functions.
\justconference{
(See~\cite{viewsizetechreport} for other tests that try to 
determine whether Cohen's \textsc{Cyclic} is better than his \textsc{General}.)
}%
The $n$-wise
independent hash alone has a strong guarantee to be beneath the theoretical
bound.  However, over the eleven Gutenberg texts, the others are just as
accurate, according to our experiments.

\subsection{Using a Wider Range of Values for $M$}

An important motivation for using $p$-wise independent hashing
is to obtain a reliable estimate while only hashing once, using a
small $M$.  Nevertheless, we have thus far not observed notable
differences between the different hash functions.  Therefore,
although we expect typical values of $M$ to a be few hundred to a
few thousand, we can broaden the range of $M$ examined.  Although
the theoretical guarantees for tiny $M$ are poor, perhaps typical
results will be usable.  And even a single buffer with $M = 2^{20}$ 
is inconsequential when a desktop computer has several gibibytes 
of RAM, and the construction of a hash table or B-tree with 
such a value of $M$ is still quite affordable.
Moreover, with a wider range of $M$, we start to see
differences between some hash functions.

\owen{5grams has the interesting ID37 problem.  Dunno about 10.
10grams would make more sense}

We choose $M=16$, $16^2$, $16^3$ and $16^4$ and analyze the 5-grams
in the text 00ws1 (Shakespeare's First Folio).
There are approximately 300,000 5-grams,
and we selected a larger file
because when $M = 16^4$ it seems unhelpful to estimate the number of
5-grams unless the file contains substantially more
5-grams than $M$.

\begin{figure}
\begin{center}
\includegraphics[width=0.5\columnwidth,angle=270]{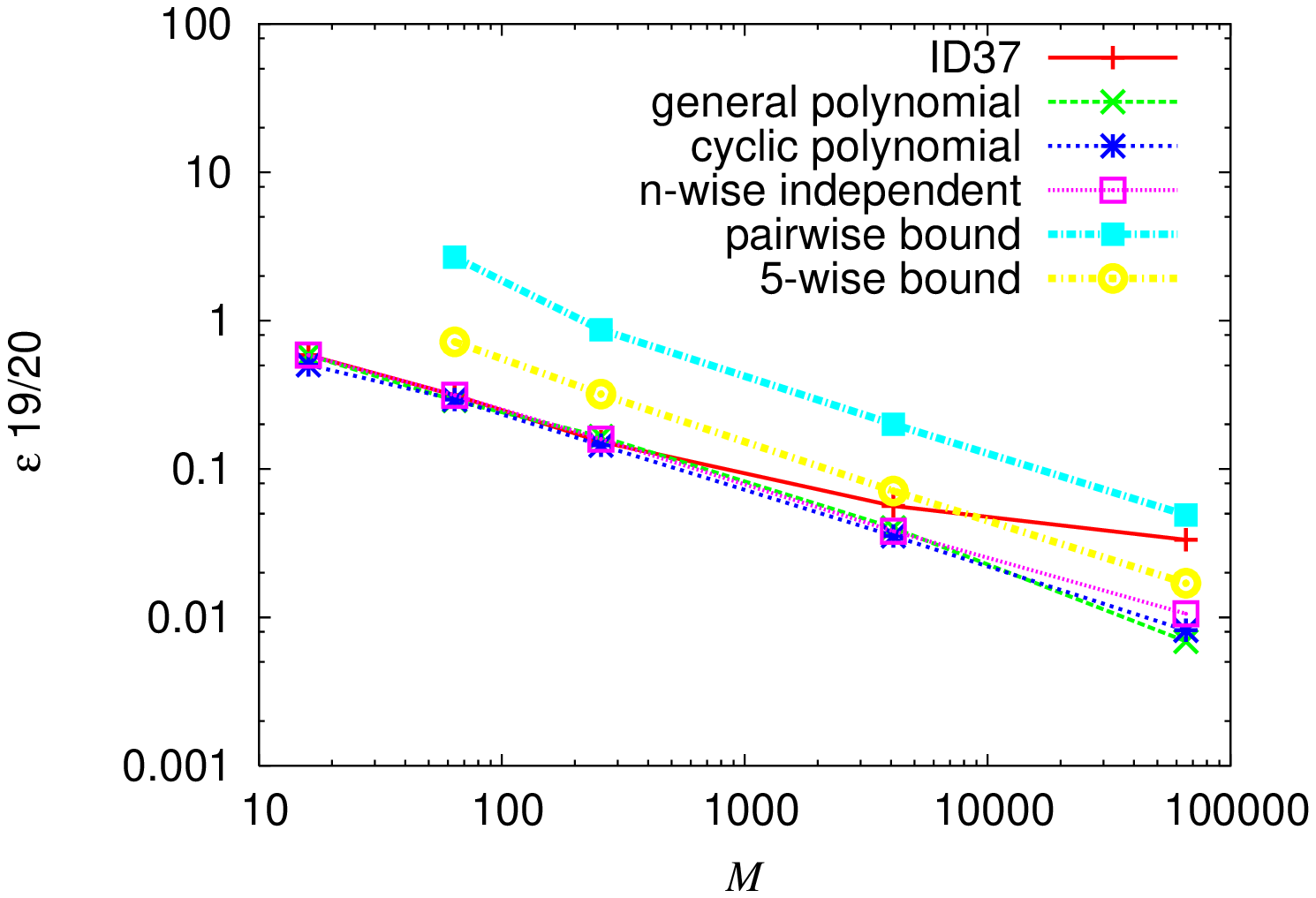}
\end{center}
\caption{\label{epsilon-vs-M-shakespeare} $95^\textrm{th}$-percentile 
error values (for 5-gram estimates) on 00ws1 for various hash families, over a wide range of $M$.
\notconference{Our analysis does not permit prediction of error bounds when $M=16$.}
 }
\end{figure}

Figure~\ref{epsilon-vs-M-shakespeare} shows the $95^\textrm{th}$-percentile 
errors for Shakespeare's First Folio, when 5-grams are estimated.
There are some smaller differences for $M = 65536$ (surprisingly, the
$5$-wise hash function, with a better theoretical guarantee, seems
to be slightly worse than Cohen's hash functions).  However, it is clear
that the theoretical deficiencies in ID37 finally have an effect: it is
small when $M = 4096$ but clear at $M = 65536$. (We observed similar
problems on Zipfian data also.) To be fair, this non-uniform
hash is still performing better than the pairwise bound, but the
trend appears  clear. 
\notconference{Does it, however, continue for very large $M$?

\subsection{Very Large $M$}

Clearly, it is only sensible to measure performance when $M \ll m$.
Therefore, we estimate the number of
$10$-grams obtained when \emph{all} plain-text files in the Gutenberg CD
are concatenated. When $M = 2^{20}$, $10$-wise independent hashing
had an observed $95^{\textrm{th}}$-percentile error of 0.182\% and
\textsc{General} had 0.218\%.  
The ID37 error was somewhat worse, at 0.286\%.  
(The theoretical pairwise
error bound is 0.908\% and the $10$-wise bound is 0.425\%.  
\owen{(If I interpret  newcomputetheoreticalestimates correctly.)})
Considering
the $M=65536$ case from Fig.~\ref{epsilon-vs-M-shakespeare}, we
see no experimental reason to prefer $n$-wise hashing to \textsc{General}, 
but ID37 looks less promising.  However, $n=10, B=37$
is a non-uniform combination for Integer Division.
}

\notconference{
\subsection{Caveats with Random-Number Generators}

To observe the effect of \emph{fully} independent hashing, we
implemented the usual (slow and memory-intensive) scheme where a 
random value is assigned and stored whenever a key is first seen.
Clearly, probabilistic counting of $n$-grams is likely to expose 
deficiencies in the random-number generator and therefore 
different techniques were tried.  The pseudo\-random-number generator
in the GNU/Linux C library was tried, as were the 
Mersenne Twister (MT)~\cite{matsumoto1998mtd} and
also the Marsaglia-Zaman-James (MZJ) 
generator~\cite{mars:rng-tr,james:pseudorandom-review,bour:uniformRNGpage}.
We also tried using a collection of bytes generated from a random physical
process (radio static)~\cite{haah:randomorg}.

For M=4096, the $95^\textrm{th}$-percentile error for text 00ws1 was
4.7\% for Linux \texttt{rand()}, 4.3\% for MT and 4.1\% for MZJ.
These three pseudorandom number generators were no match for truly
random numbers, where the 95\% percentile error was only 2.9\%.
Comparing this final number to Fig.~\ref{epsilon-vs-M-shakespeare},
we see fully independent hashing is only a modest improvement on
Cohen's hash functions (which fare better than 5\%) 
despite its stronger theoretical guarantee. 

The other hash functions also rely on random-number generation
(for $h_1$ in Cohen's hashes and ID37; for $h_1 \ldots h_n$ in the
$n$-wise independent hash).  It would be problematic if their performance
were heavily affected by the precise random-number generation process.
However, when we examined the $95^\textrm{th}$-percentile errors
we did not observe any appreciable differences from varying the
pseudo\-random-number generation process or using truly random
numbers. 
(The graphs are shown in Appendix~\ref{data-appendix}.)
Surprisingly, the pseudorandom generators may have been marginally
\emph{better} than the truly random numbers.
}

\notconference{
\subsection{$\textrm{95}^\textrm{th}$-Percentile Errors Using Zipfian Data}

\begin{figure}
\begin{center}
\includegraphics[height=0.45\columnwidth,angle=270]{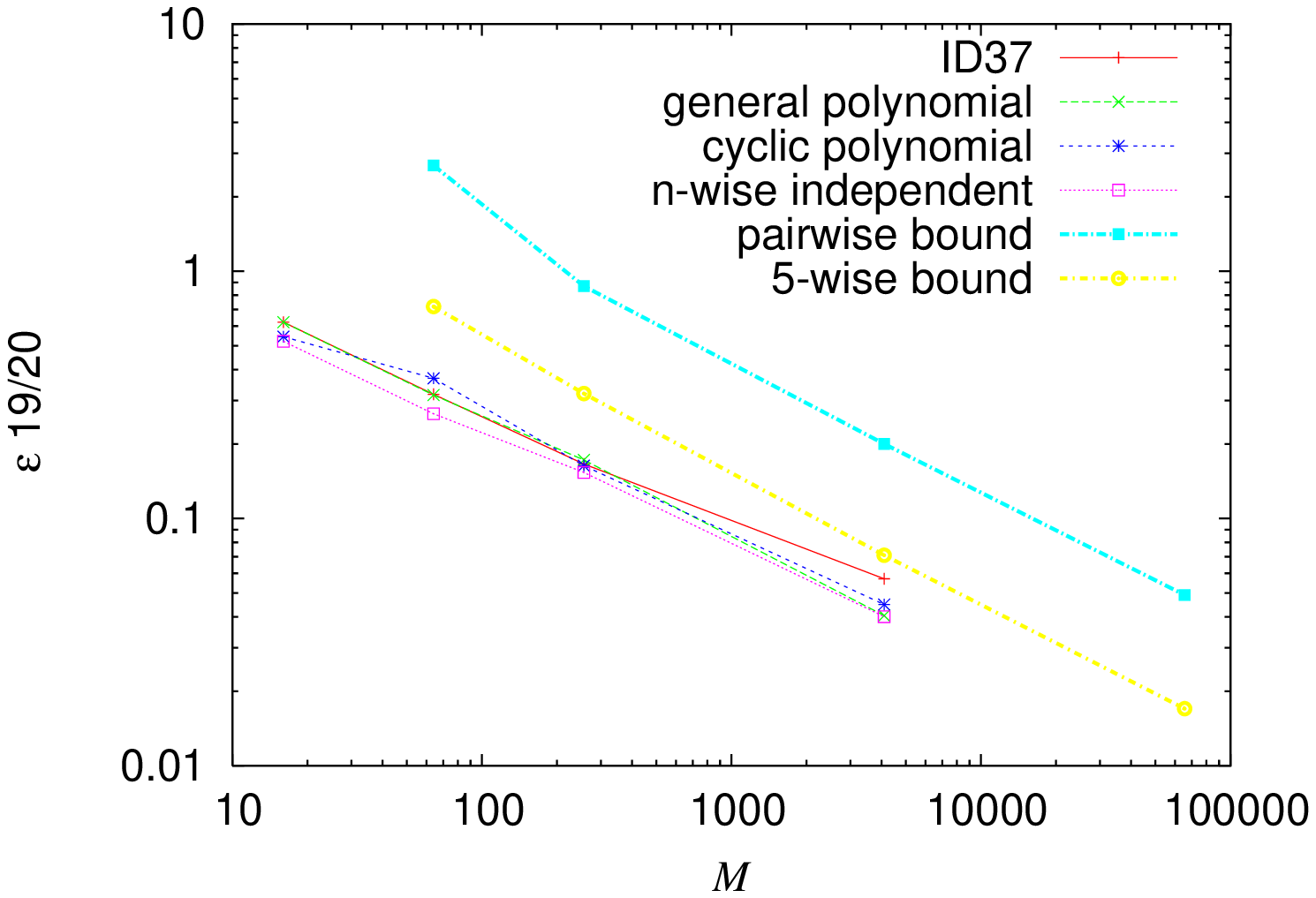}
\includegraphics[height=0.45\columnwidth,angle=270]{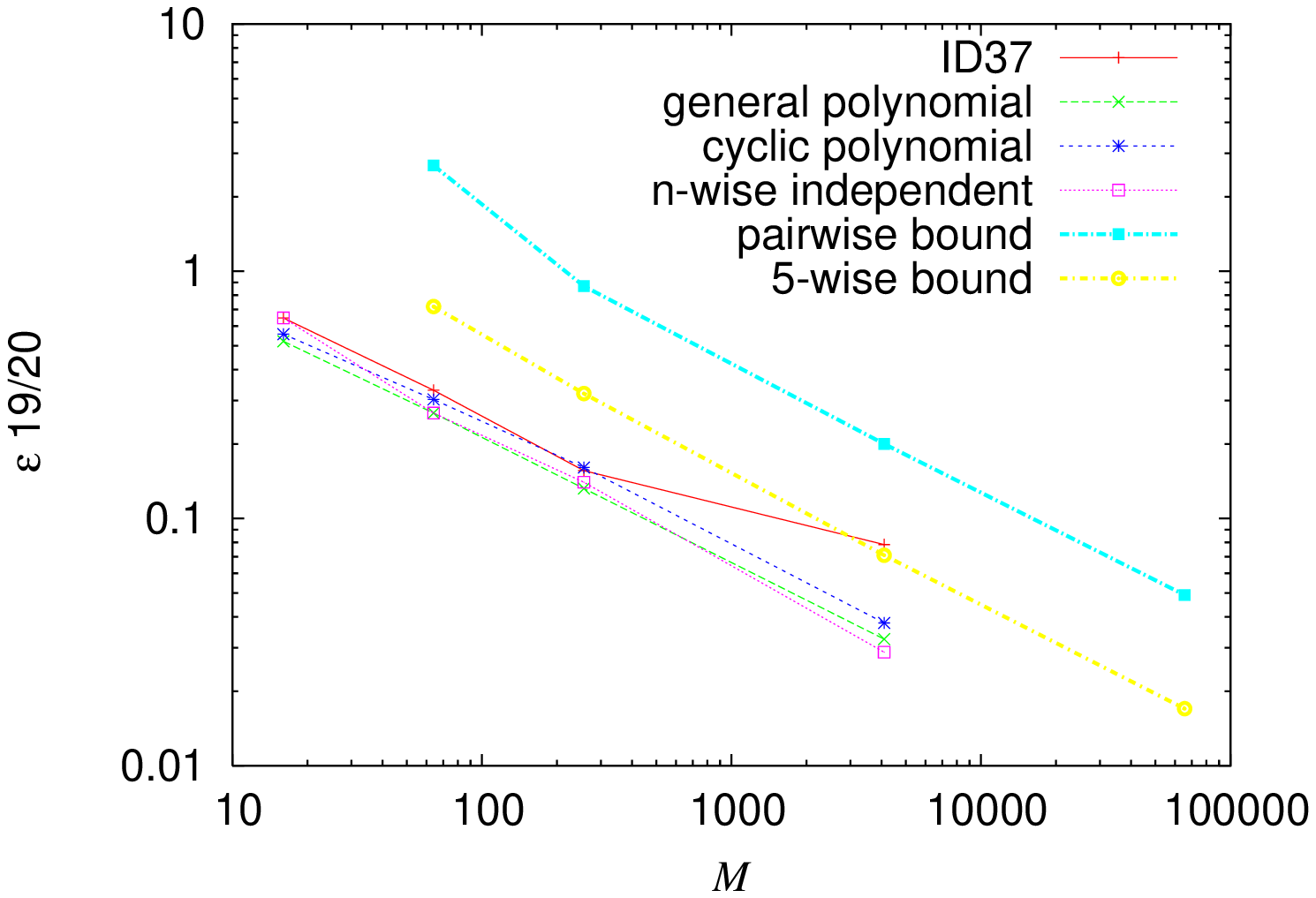}\\
\includegraphics[height=0.45\columnwidth,angle=270]{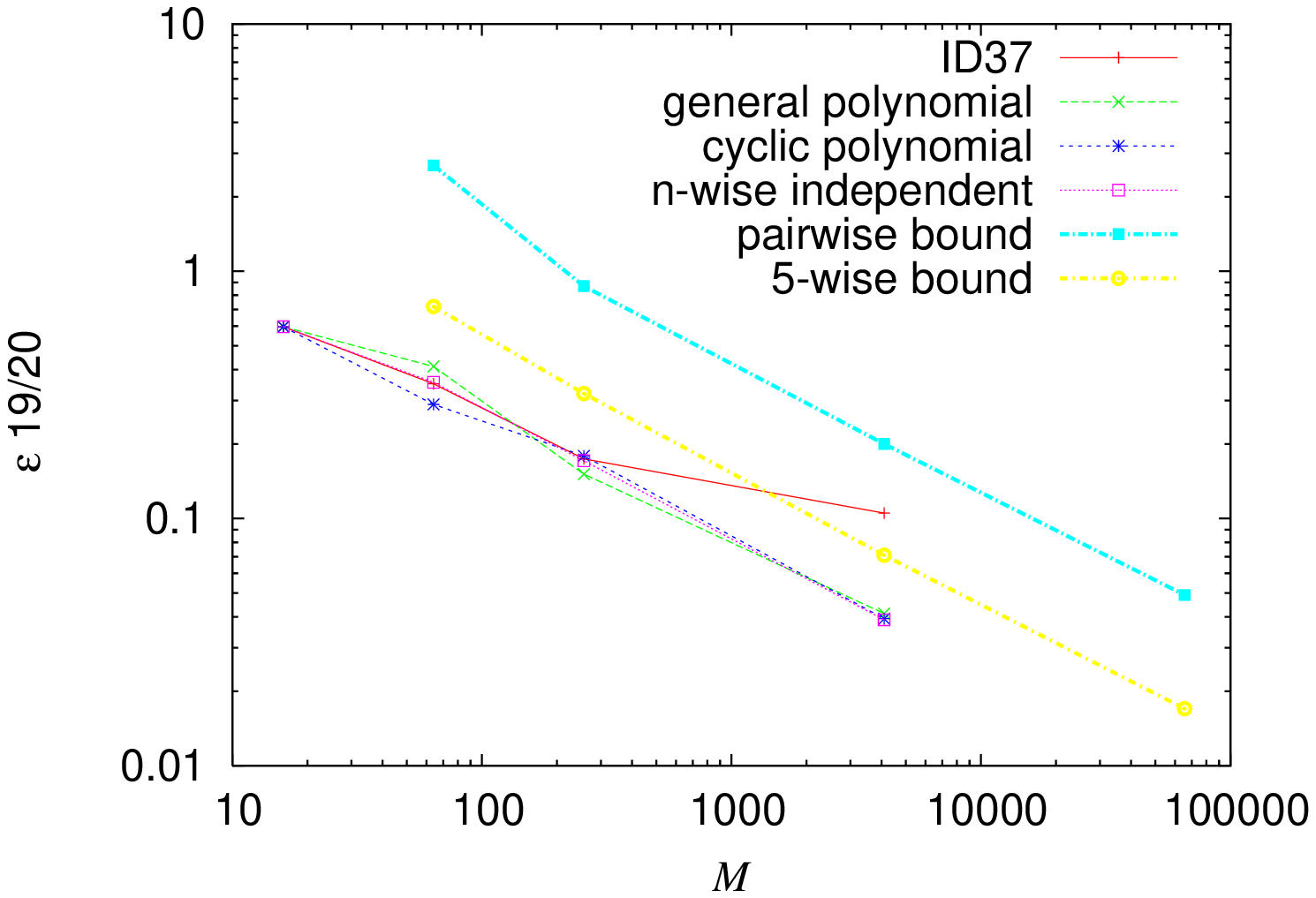}
\includegraphics[height=0.45\columnwidth,angle=270]{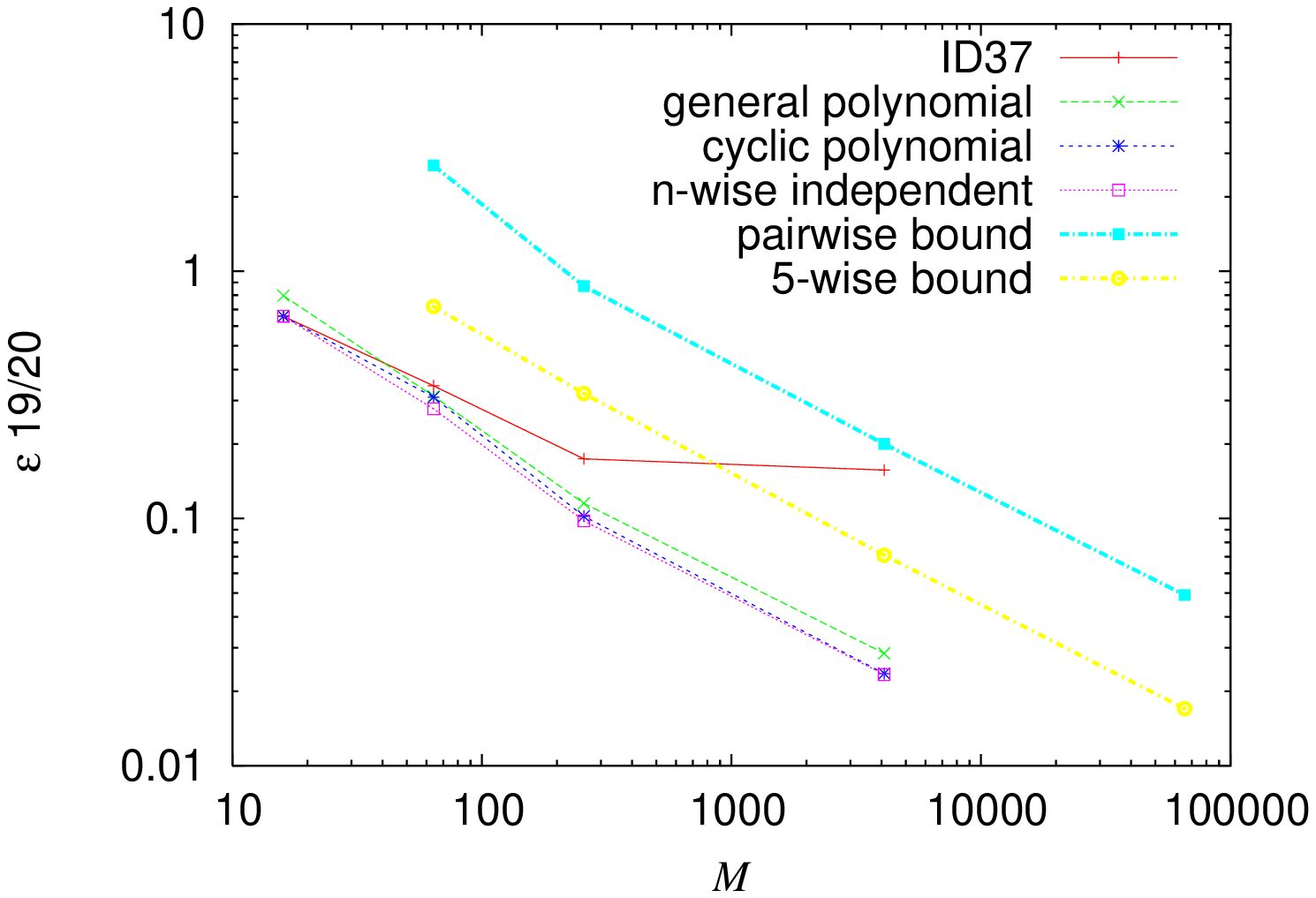}
\end{center}
\caption{\label{zipf-five}n=5 Zipfian data,  $s=1$ (top left),
$s=1.2$ (top right), $s=1.6$ (bottom left) and $s=2$ (bottom right).}
\end{figure}

The various hash functions were also tested on synthetic Zipfian data, 
where the probability of the $k^\textrm{th}$ symbol is
proportional to $k^{-s}$.  
(We chose $s \in [0.8, 2.0]$.)
\komment{ 0.8, 1.0, 1.2, 1.4, 1.6, 1.8 and 2.0.}  
Each data set had $N \approx 10^5$, but
for larger values of $s$ there were significantly fewer $n$-grams.  Therefore,
measurements for $M$=65536 would not be meaningful and are omitted.


Some results, for $n=5$, are shown in Fig.~\ref{zipf-five}.
The ID37 method is noteworthy.  Its performance for larger $M$
is badly affected by increasing $s$.
The other hash functions are nearly indistinguishable in almost all other cases.
Results are similar for 5-grams and 10-grams (which are not shown), 
except that $s$ can grow slightly bigger  for 10-grams before ID37 
fails badly.  

\komment{
\begin{figure}
\begin{center}
\end{center}
\caption{\label{zipf-five-two}n=5 Zipfian data, s=1.6 (left)  s=2 (right)}
\end{figure}
}

\subsection{Choosing Between Cohen's Hash Functions}

The experiments reported so far do not reveal a clear distinction,
at least at the $95^\textrm{th}$ percentile error, between
Cohen's \textsc{General} and his \textsc{Cyclic} polynomial hashes.  
This is somewhat
surprising, considering the theoretical differences (nonuniformity
versus pairwise independence)  shown
in Lemma~\ref{not-uniformlemma} 
and Lemma~\ref{uniformlemma}.

The ratio $m/M$ may be significant (because
this ratio affects how many hash bits are used) and
our experiments thus cover two different ratios.
We first used synthetic
Zipfian data ($s=2$), looking for $5$-grams, since
that combination revealed the weakness of ID37.  Since 
$m=14826$ for our Zipfian data set, choosing $M=64$ 
gives a ratio about 231 and $M=1024$ gives a ratio of about 14.

Results, for more than 2000 runs, are shown in 
Table~\ref{zipf-competition}. 
The top of the table shows $\epsilon$ values (percents), with boldfacing 
indicating a case when one technique had a lower error than the other.
It shows a \emph{slightly} better  performance for \textsc{Cyclic}.
Means also slightly favour \textsc{Cyclic}. This is consistent with
the experimental results reported by Cohen~\cite{cohenhash}.

\begin{table}[H]
\caption{\label{zipf-competition} Comparing polynomial hashes
\textsc{Cyclic} and \textsc{General}, Zipfian data set.}
\begin{center}
\begin{tabular}{|c||ll|ll|} \hline
percentile & \multicolumn{2}{c|}{\textsc{Cyclic}} & 
             \multicolumn{2}{c|}{\textsc{General}}\\
   & M=64              & M=1024          & M=64  & M=1024\\ \hline
25 &  3.60             & 0.931           &   3.60&  0.931  \\
50 &  \textbf{7.06}    & \textbf{1.98}   &   8.49&  2.01   \\
75 & 14.0              & \textbf{3.41}   &  \textbf{13.7} &  3.60   \\
95 & \textbf{30.9}     & \textbf{6.11}   &  31.2 &  6.22   \\ \hline
mean &  \textbf{10.6}  & \textbf{2.45}   &  10.7 &  2.51   \\ \hline
\end{tabular}
\end{center}
\end{table}

We also ran more extensive tests using 00ws1, where there seems 
to be no notable distinction. 
10,000 test runs were made.  
Results are shown in Table~\ref{shakes-competition-cohen}.
The overall conclusion is that the theoretical advantage held by
\textsc{General} does not carry over.  Experimentally,
these two techniques cannot be distinguished meaningfully 
by our tests.
}

\begin{table}[H]
\caption{\label{shakes-competition-cohen} Comparing polynomial hashes
\textsc{Cyclic} and \textsc{General}, data set 00ws1}
\begin{center}
\begin{tabular}{|c||ll|ll|} \hline
percentile & \multicolumn{2}{c|}{\textsc{Cyclic}} & 
             \multicolumn{2}{c|}{\textsc{General}}\\
   & M=64              & M=1024          & M=64  & M=1024\\ \hline
25 &  5.79             & 1.30            &  5.79 &  1.30  \\
50 &  10.7             & \textbf{2.58}   &  10.7 &  2.69   \\
75 & \textbf{18.2}     & 4.55            &  19.0 &  4.55   \\
95 & 30.6              & 7.69            &  30.6 &  7.69   \\ \hline
mean &  \textbf{12.5}  & 3.15 &  12.6 & \textbf{3.14}   \\ \hline
\end{tabular}
\end{center}
\end{table}

\komment{  not going to do experimentally for this conf paper}
\subsection{Estimating Iceberg Counts and Entropy}

\begin{figure}[tb]
\begin{center}
\subfigure[Entropy estimates]{
\includegraphics[height=0.55\columnwidth,angle=270]{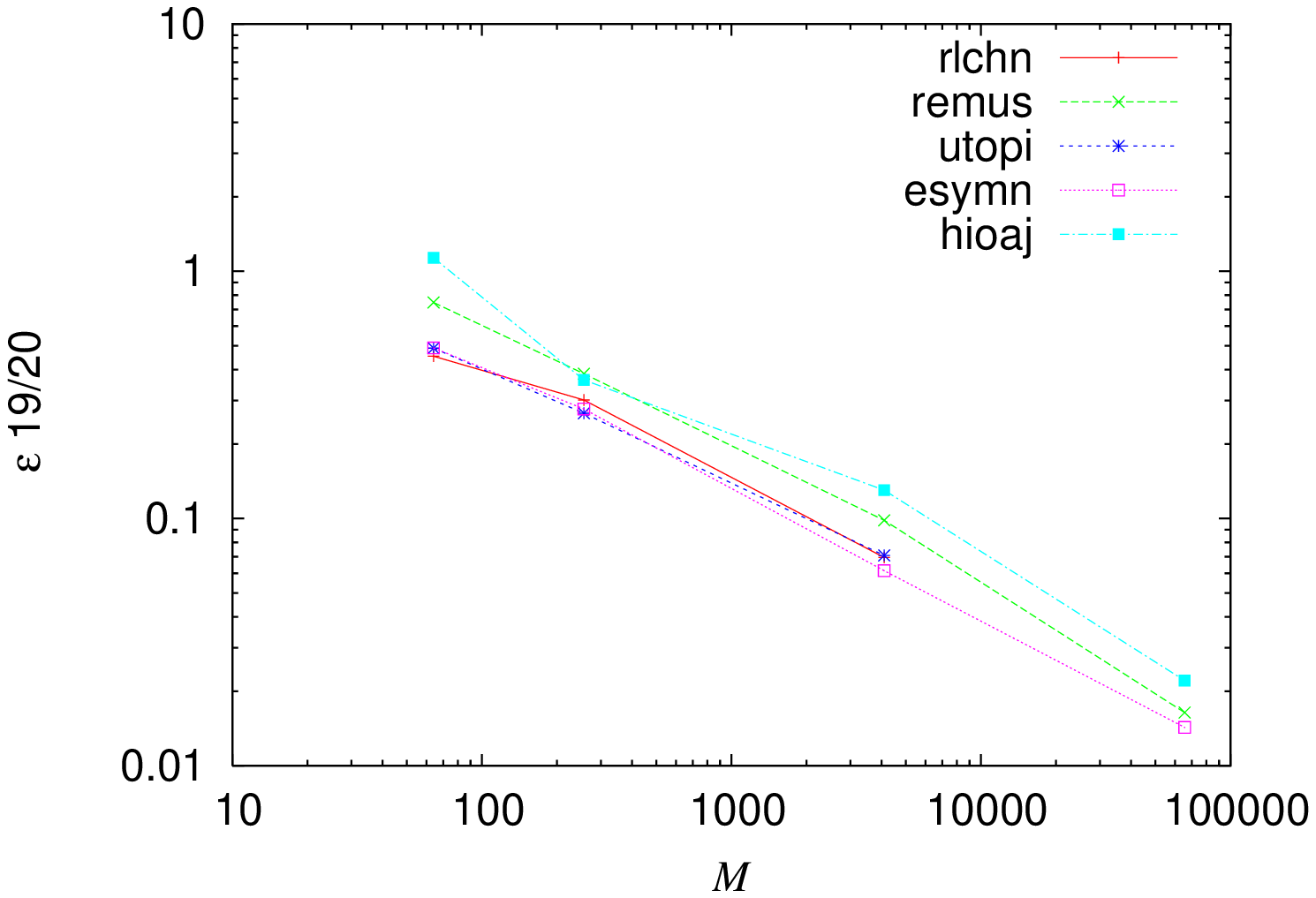}
}
\subfigure[Iceberg-count estimates]{
\includegraphics[height=0.55\columnwidth,angle=270]{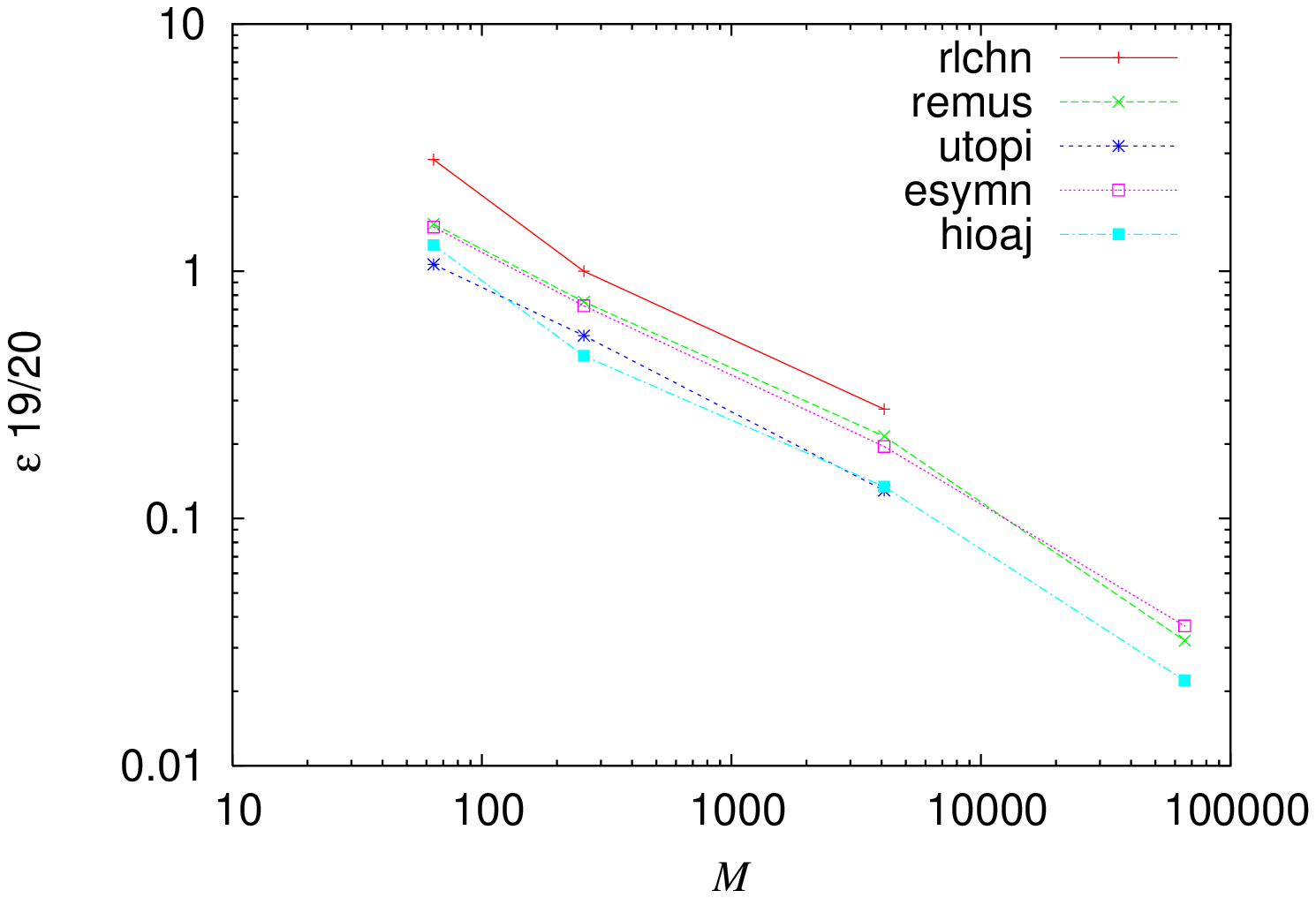}
}
\end{center}
\caption{\label{ent-ice-error-small}Relative errors 
on smaller data sets: rlchn, remus, utopi, esymn and hioaj.
Points with zero error are omitted, due to the logarithmic axes.}
\end{figure}

\begin{figure}[tb]
\begin{center}
\subfigure[Entropy estimates]{
\includegraphics[height=0.55\columnwidth,angle=270]{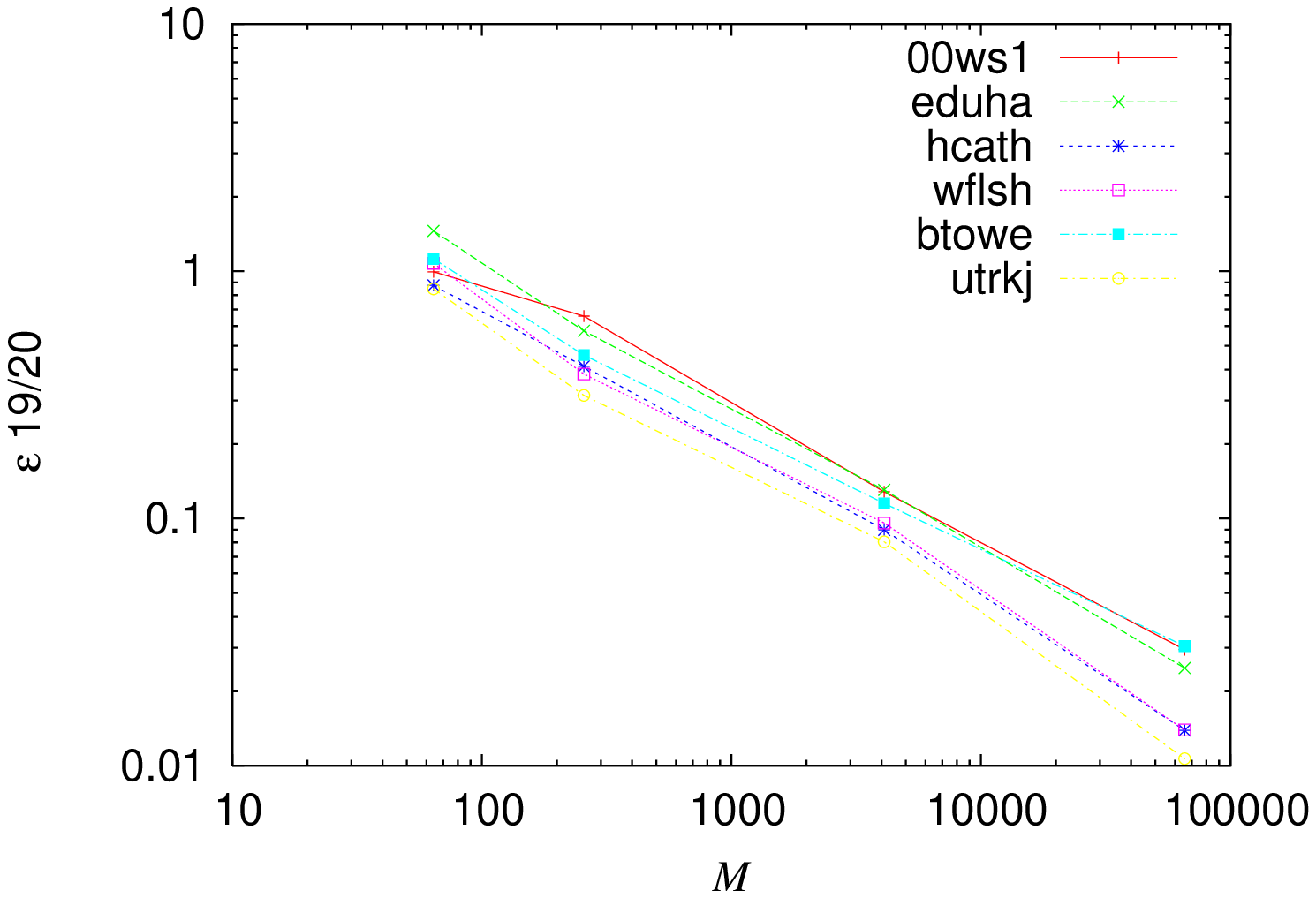}
}
\subfigure[Iceberg-count estimates]{
\includegraphics[height=0.55\columnwidth,angle=270]{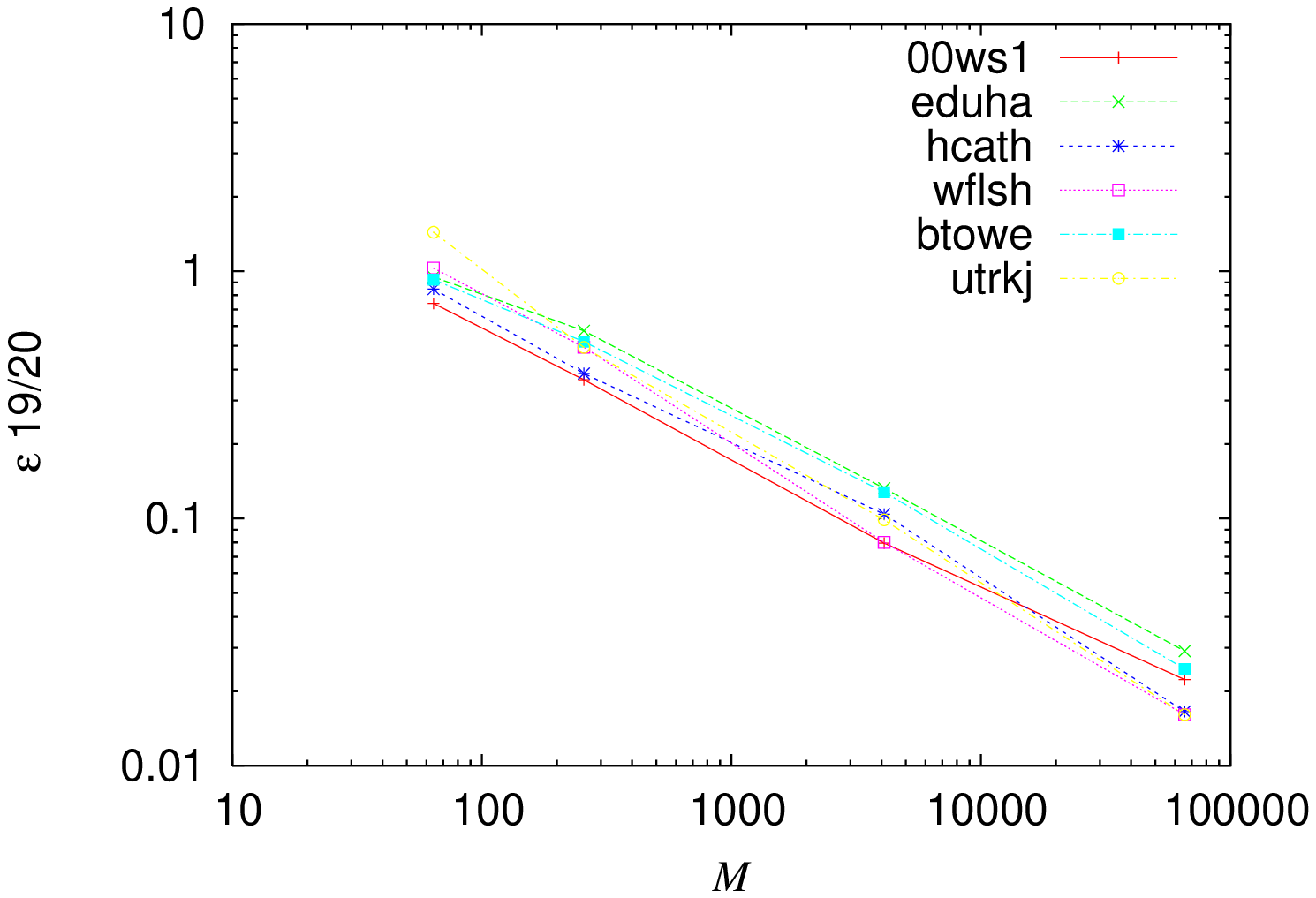}
}
\end{center}
\caption{\label{ent-ice-error-big}Relative errors 
on larger data sets: 00ws1, eduha, hcath, wflsh, btowe and  utrkj.}
\end{figure}

Although we do not have a theoretical bound to compare with, experiments
tested
the approach to single-pass iceberg-count and entropy estimation given Section~\ref{iceberg-entropy}.  On our 11 data sets, the
amounts of relative error (for 5-grams) observed with a 95\% reliability are
shown in Figs.~\ref{ent-ice-error-small}--\ref{ent-ice-error-big}.
(The data is shown separately for the 6 data sets with 100,000 or more
5-grams, since the $M=65536$ case is uninteresting for the other 5 data sets.)
The iceberg-count estimates the number of distinct $n$-grams occurring
at least 10 times.

We see that the error of estimates does decrease quickly with $M$, and that
when $M < 4096$, the accuracy was poor.  However, the accuracy when $M = 4096$
might be adequate for some applications.

\komment{

\begin{figure}
\begin{center}
\includegraphics[width=0.85\columnwidth]{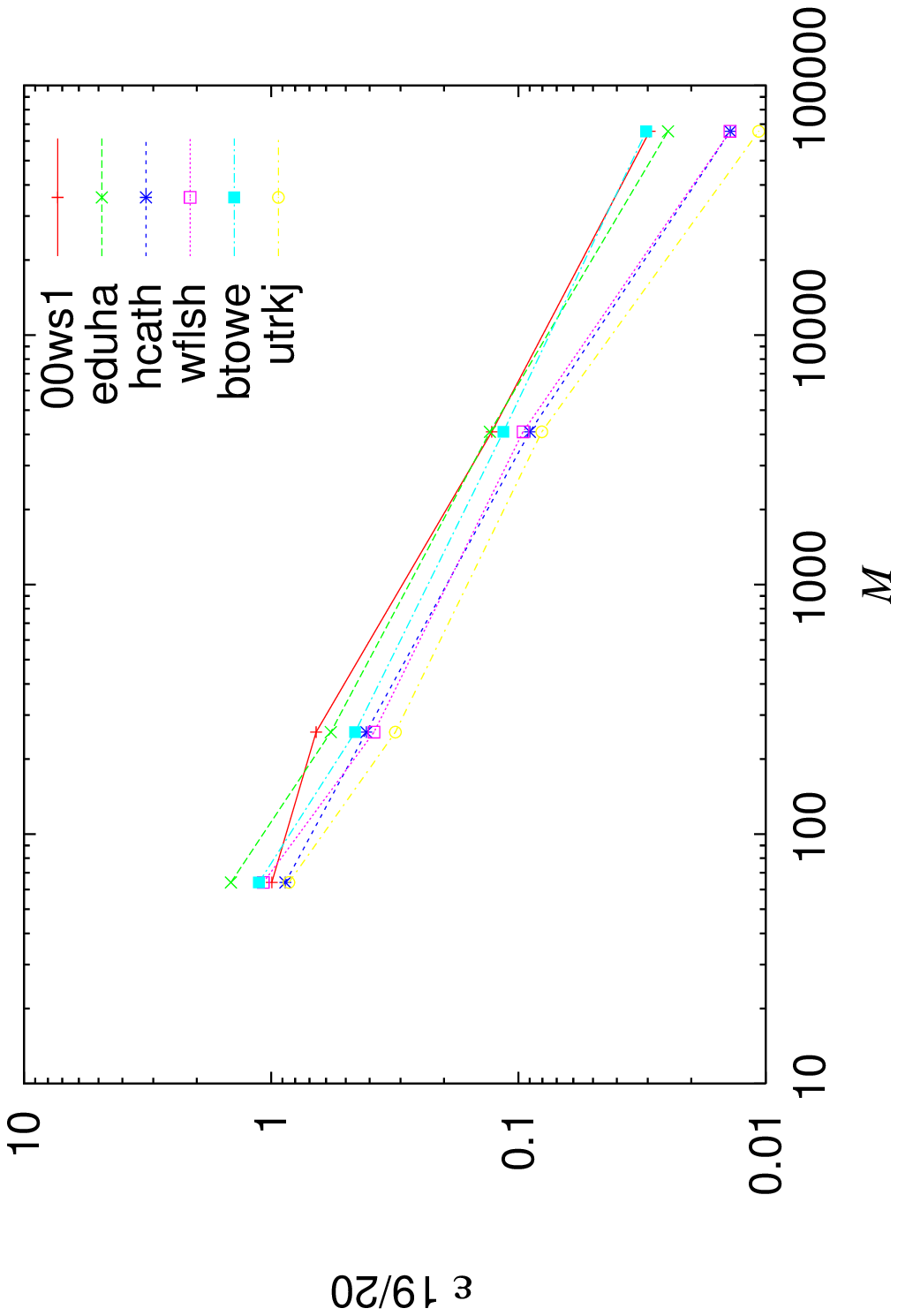}
\end{center}
\caption{\label{ent-error-second}Relative errors for entropy estimates, on utopi, wflsh, btowe and esymn}
\end{figure}

\begin{figure}
\begin{center}
\includegraphics[width=0.85\columnwidth]{epsilon-vs-M-n5-entropy3.eps}
\end{center}
\caption{\label{ent-error-third}Relative errors for entropy on hioaj, rlchn and utrkj.}
\end{figure}
}

\komment{

\begin{figure}
\begin{center}
\includegraphics[width=0.85\columnwidth]{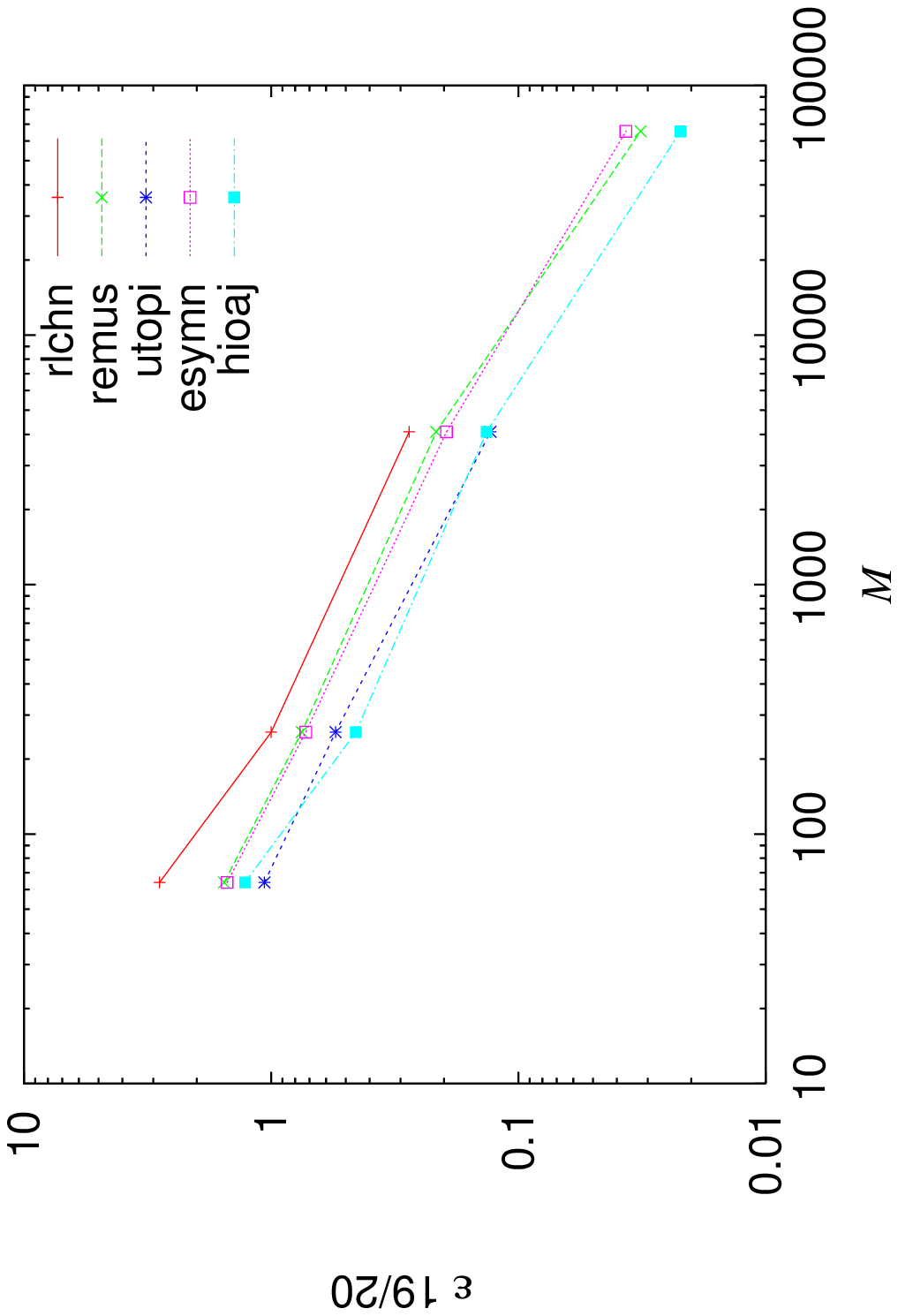}
\end{center}
\caption{\label{ice-error-first}Relative errors for iceberg-count
(10 occurrences or more) estimates
on 00ws1, eduha, hcath and remus.
Points with zero error are omitted, due to the logarithmic axes.}
\end{figure}

\begin{figure}
\begin{center}
\includegraphics[width=0.85\columnwidth]{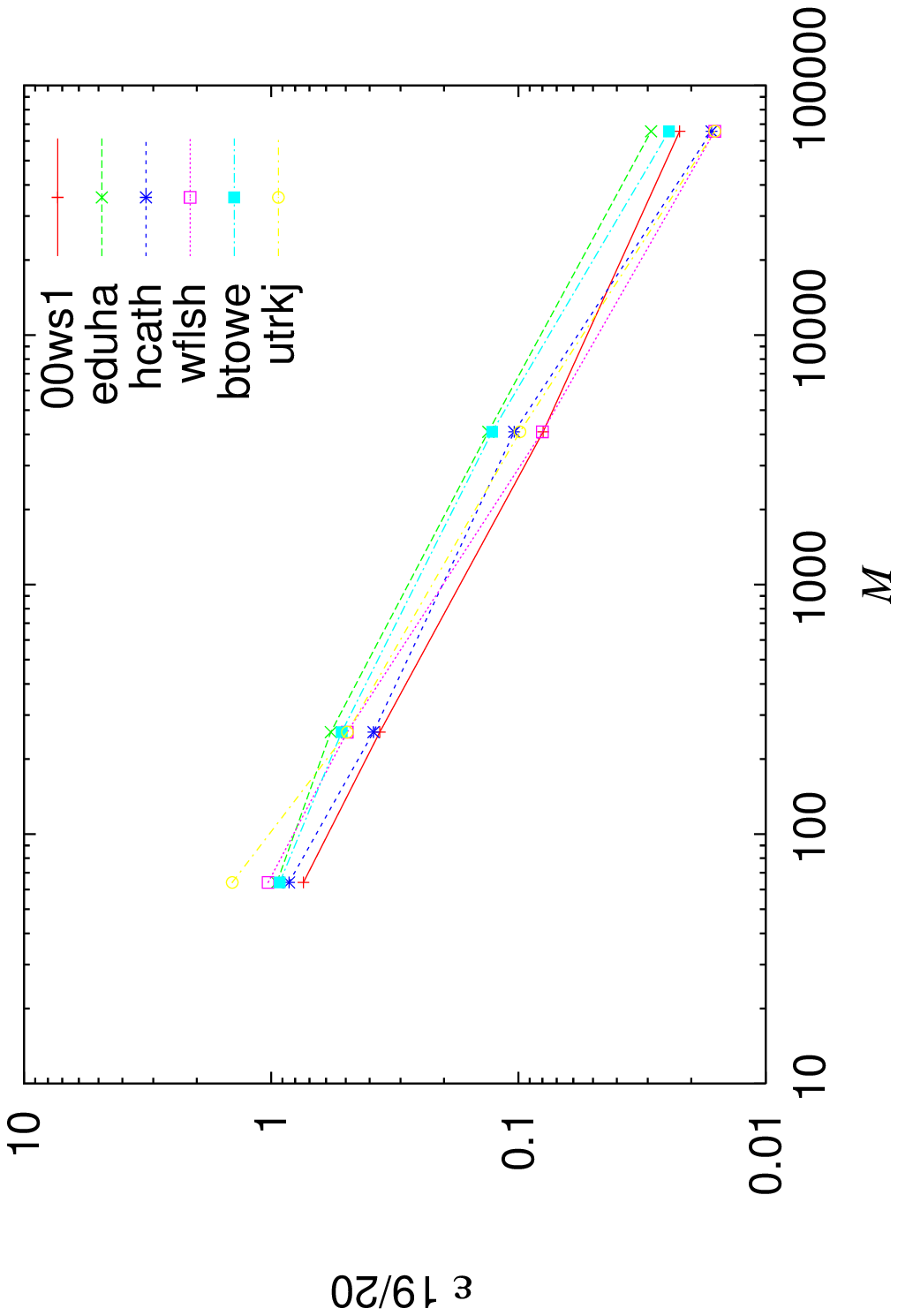}
\end{center}
\caption{\label{ice-error-second}Relative errors for iceberg-count estimates, on utopi, wflsh, btowe and esymn}
\end{figure}

\begin{figure}
\begin{center}
\includegraphics[width=0.85\columnwidth]{epsilon-vs-M-n5-iceberg3.eps}
\end{center}
\caption{\label{ice-error-third}Relative errors for iceberg-count on hioaj, rlchn and utrkj.}
\end{figure}
}

\komment{
 Due to the small sizes of some test
inputs, $M = 65536$ is uninteresting for inputs 
rlchn, remus, utopi, esymn and hioaj 
(which each have 100,000 or fewer 5-grams).
}

While we expected poor results on English texts due to the biased distribution,
the results suggest that useful theoretical bounds are possible.

\subsection{Speed}

Speeds were measured on a
Dell Power Edge 6400 multiprocessor server 
(with four Pentium III Xeon 700\,MHz processors having 2\,MiB cache each,
sharing 2\,GiB of 133\,MHz RAM).  The OS kernel was Linux~2.4.20 and 
the GNU C++
compiler version~3.2.2 was used with relevant compiler flags 
\texttt{-O2 -march=i686  -fexceptions}. The STL \textit{map}
class was used to construct look-up tables.

Only one processor was used, and the data set
consisted of all the plain text files on the Project Gutenberg CD,
concatenated into a single disk file containing
over 400\,MiB and approximately 116~million 10-grams. 
For comparison, this file was too large to process with the Sary suffix 
array~\cite{sary-home} package (version 1.2.0), since the array would have
exceeded 2\,GiB.
However, the first 200\,MB 
\emph{was} successfully processed by Sary, which took 1886\,s to build the 
suffix\notconference{\footnote{
Various command-line options were attempted and the
reported time is the fastest achieved.}
\footnote{Pipelined suffix-array implementations reportedly can
process inputs as large as 4~GB in hours~\cite{dementiev2005bem}.}}
array.
\notconference{The SUFARY~\cite{sufary-home} (version 2.3.8) package is said to be 
faster than sary~\cite{sary-home}.  It processed the 200\,MB file
in 2640\,s and then required 95\,s to (exactly) compute the number
of 5-grams with more than 100,000 occurrences.}


\begin{table}
\caption{\label{speedtable}Time (seconds) to process all Gutenberg CD files, $10$-grams.} 
\begin{center}
\begin{tabular}{|l|r|r|}\hline
Hashing    & $M  = 2^{10}$ & $M  = 2^{20}$ \\ \hline
$10$-wise  & 794           & 938   \\
ID37       & 268           & 407   \\
Cyclic     & 345           & 486   \\
General    & 352           & 489   \\ \hline
\end{tabular}
\end{center}
\end{table}

From Table~\ref{speedtable} we see that $n$-gram estimation can be efficiently
implemented.  First, comparing results for $M=2^{20}$ to those for 
$M=2^{10}$, we see using a larger table costs roughly 140\,s in every
case. This increase is small when considering that $M$ was multiplied
by $2^{10}$ and is consistent with the fact that the computational
cost is dominated by the hashing.  Comparing different hashes,
using a $10$-wise independent hash was about twice as slow
as using a recursive hash.  Hashing with ID37 was 15--25\% faster than
using Cohen's approaches.

Assuming that we are willing to allocate very large files to create
suffix arrays and use much internal memory, an exact count is
still at least 10 times more expensive than an approximation.
Whereas the suffix-array approach would take about an hour to compute
$n$-gram counts over the entire Gutenberg~CD, an estimate can
be available in about 6\,minutes while using very little memory
and no permanent storage.

\komment{
\section{Entropy Estimation by Probing}

Notational issue, don't use $p_i$ ideally, to avoid confusion
with $p(x)$.

\owen{I think it (and iceberg cubes) deserve mention and some quick
experiments. If others have used GT on entropy, then I would not bother,
though.}

The entropy rate is defined as $\sum_i p_i \log p_i$.

*) when estimating the entropy, if you use a random string, then all
 $p_i$'s are the same or almost so... hence $p_i \log p_i$ is constant, and
 the entropy rate is $1/n \log 1/n$ where n is the number of distinct
 symbols. This means that for a random string, estimating the entropy
 or estimating the count is the same thing. (Do you grok this? I can
 explain better.)

*) For other types of strings, $p_i \log p_i$ can vary wildly (the n-gram
 "the" is much more likely than others). Suresh' paper discuss this:
 how you have to work a bit harder to estimate the entropy when it
 gets lower. So it makes sense that entropy estimation would be
 sensitive to entropy (the less entropy, the harder it is to estimate
 it).

Shannon wrote that the entropy of $n$-grams for the English language
would be between  0.6 and 1.3 as $n$ grows~\cite{NagaoMori} but such 
a computation would require a virtually infinite English text.

Talk about marginal entropy bounds. Given the entropy of a matrix
(each entry of the matrix being a count value) you can bound it by 
the marginal entropies (see entropy.py script). You can apply this
to n-grams to say that entropy(n-grams)<= n* entropy(1-grams). 
(Application of these marginal entropy estimation results
assume a circular string.)

In a data cube setting, let C be the big cube, and let $C_i$  be
the $k$-dimensional cuboids, then $k*entropy(C) \leq\sum entropy(C_i)$.
The k in k*entropy(C) comes from the fact that each attribute value is
repeated k times on the right.

Also that the k*entropy(n-grams)<= n * entropy (k-grams).

Possibly, our good friend Shannon derived these things? I don't have
a reference for these things, but they are related to the geometry problem
(you have n points in space, and $n_x$ point in one projection 
where you omitted
the x axis, and so on, then you can prove that$ n^2 \leq n_x * n_y * n_z$

Conjecture: most English texts have some value of n where the
number of unique n-grams  is roughly L/2.  And this value, for the
particular English text, says something significant about it...

At this point the *entropy* of the n-grams is really, really
high. This means that if you based a random text generator one k-grams
statistics for k<n, it would pass as valid English.

Interestingly, this is entirely general. For example, you could
simulate stock market data (but that will be for another time) using
this sort of techniques.

(Tough to estimate the entropy of a language with finite texts)
}
\section{Conclusion}

Considering speed, theoretical guarantees, and actual results, we
recommend Cohen's \textsc{General}.  It is fast, has a theoretical performance guarantee,
and behaves at least as well as either ID37 or the  $n$-wise independent
approach. \textsc{General} is pairwise independent so that there are minimal
theoretical bounds to its performance. The $n$-wise independent hashing 
comes with a stronger
theoretical guarantee, and thus there can be no unpleasant surprises
with its accuracy on any data set.
Yet there is a significant
speed penalty for its use in our implementation.
The speed gain of ID37 
is worthwhile only for very small values of $M$.  Not only does it lack
a theoretical accuracy guarantee, but for larger $M$ it is observed to
fall far behind the other hashing approaches in practice.  Except where 
accuracy is far less important than speed, we cannot recommend ID37. 

Iceberg-count and entropy estimates on English text showed 
good decay with larger values of $M$ warranting further theoretical
investigations.


\komment{

What about ``hashing'' using library implementations of things like CRC
and the other checksummers (eg as in the Java library).   Again, I would
want to know if I were a practical person.
\owen{invoking Java from python would require jython; if from C++
I am not sure how to do it. Postponed.}

\daniel{AH! Easily done in Python:}

\begin{verbatim}
> import zlib
> zlib.crc32("fdsfsd")
2025892861
> zlib.adler32("fdsfsd")
145097339
\end{verbatim}

\owen{Thanks.  Not sure whether it's worthwhile to do.}
}

\owen{maybe check with zlib's adler32 and crc32  from C++}

\komment{
\notconference{
\section{Other Implementation Issues}

some intro text here.

}
}

There are various avenues for follow-up work that we
are pursuing. 
\notconference{Further improvements to the theoretical
bound seem possible, especially for larger values of $p$.}
\justconference{
First, further improvements to the theoretical
bound seem possible, especially for larger values of $p$.
Second, each item being counted can have an occurrence
count kept for it.   This would enable entropy estimation as
well as estimates of the number of distinct \emph{frequent}
$n$-grams. 
}%
A more sophisticated solution to the simultaneous estimation
problem may be possible, and it could be experimentally evaluated
against the approach sketched in Section~\ref{simultaneous-section}
and against suffix-array methods.
\notconference{

Efficient frequent string mining have been recently
proposed~\cite{fisher05} to find frequent substrings in one string
that are rare in another.  However, suffix arrays do not support fast
search for the most frequent phrases containing a given word. While
suffix arrays allow us to count occurrences of a substring, they provide
no means to count the occurrences of the various $n$-grams beginning
with a given substring, let alone the $n$-grams containing a
substring.

\komment{
Experiments suggest that much stronger theoretical bounds might be
possible especially for 
\textsc{Recursive Hashing by General
Polynomials} which effectively behaves as an $n$-wise independent hash
function while not being even pairwise independent. 

\textbf{compare to most recent thy}
}

\komment{

owen:   cut since this is more datacube/cubeoid rather than ngram

Possible research track (Owen's insight): massive sharing in hashing

Also, you're working on many different cube estimates simultaneously.
Consider base-cube ABCDEFG with tuple t. We will project it onto
ABCDEF and ABCDEG and ..... and hash each projection.

Most of the work that you do in hashing "t projected onto ABCDEF"
is common to hashing "t projected onto ABCDEG", assuming that you hash
in the most obvious way. So we should be able to do massive sharing in the
hash computation. Do the Panda people try this?

In the ngram context, this is like ``ngram with subsets of the n 
positions''.

}

\komment{
Insight: "(symmetrical) joins are tight upper bounds on the number of n-grams"

Explanation: suppose I want a bound on how many trigrams there are
starting with "A B", well, that's just the number of bigrams starting
with B! Similarly, if I want to bound the number of bigrams starting
with B, I can just use the total number of unigrams!

Insight: I can use knowledge of how n-1-grams there are starting in a
certain way, to compute bounds on the number of n-grams, this is
recursive!

\owen{Is there any divide and conquer advantage to
thinking about an ngram being an n/2-gram concatenated with another
n/2 gram?}\daniel{Ok. As for divide and conquer, I don't know.}

Future work: lazy and simultaneous n-gram statistics
}
}

\bibliographystyle{alpha}
\bibliography{lemur}        

\appendix
\section{Pseudorandom-Number Generators and Accuracy}
\label{data-appendix}

The following graphs 
show that varying the source of random (or pseudorandom)
numbers had little effect on \textsc{General}, \textsc{Cyclic} or the
$n$-wise random hashes.


\begin{figure}[H]
{\centering
\includegraphics[height=0.9\columnwidth,angle=270]{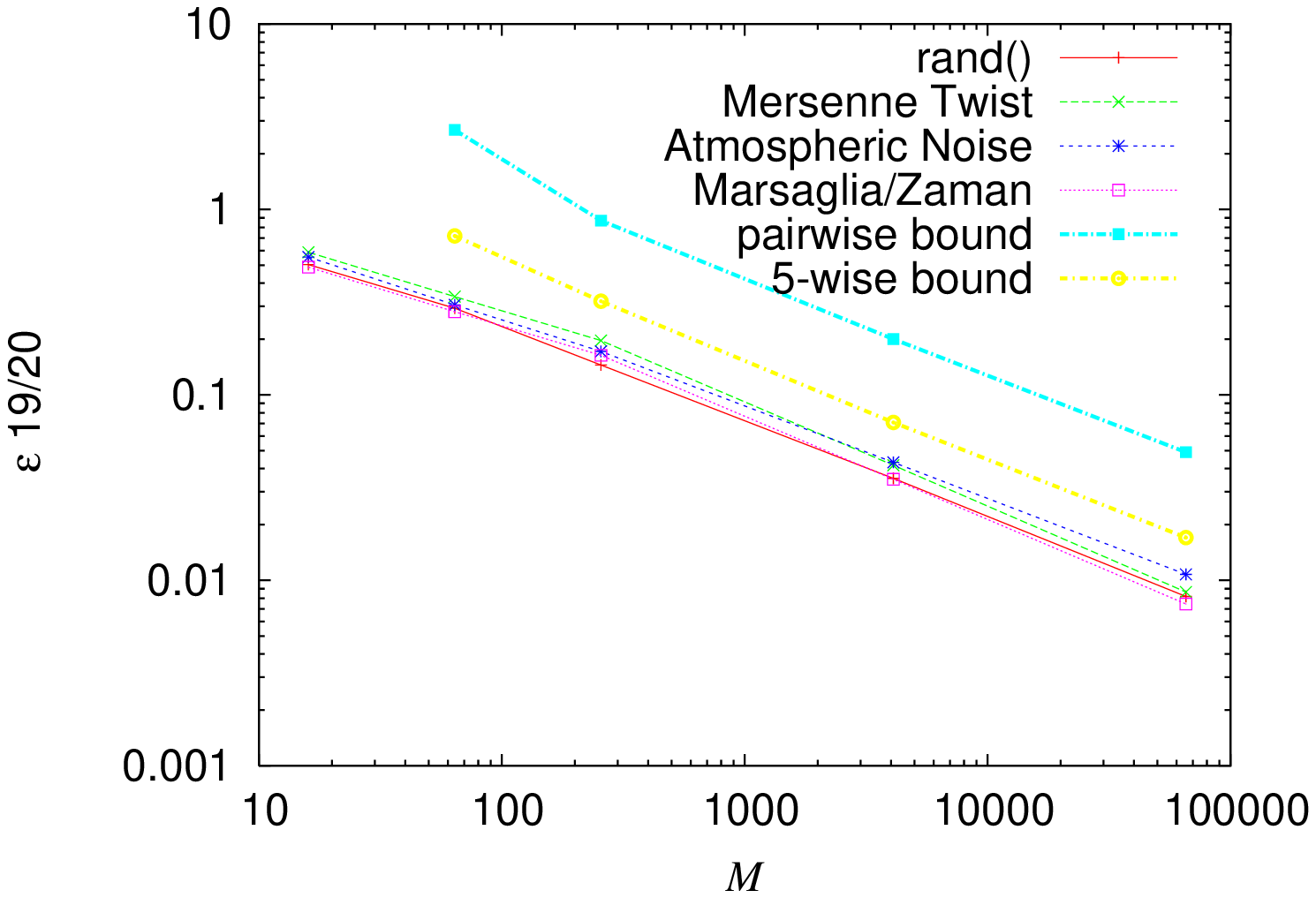}
}
\caption{\textsc{Cyclic} is not affected much
by the source of random numbers.}
\end{figure}

\begin{figure}[H]
{\centering
\includegraphics[height=0.9\columnwidth,angle=270]{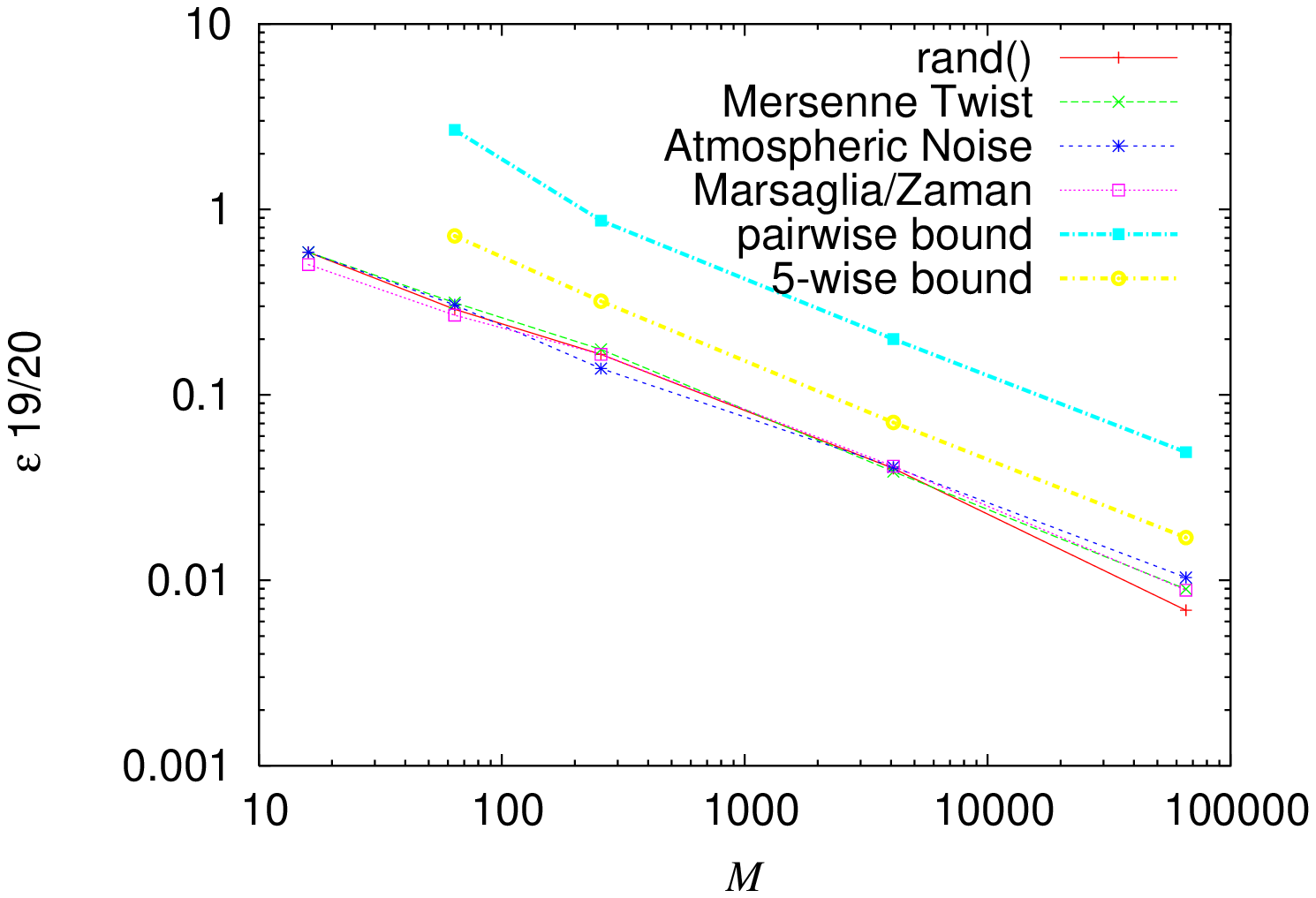}
}
\caption{\textsc{General} is similarly unaffected.}
\end{figure}

\begin{figure}[H]
{\centering
\includegraphics[height=0.9\columnwidth,angle=270]{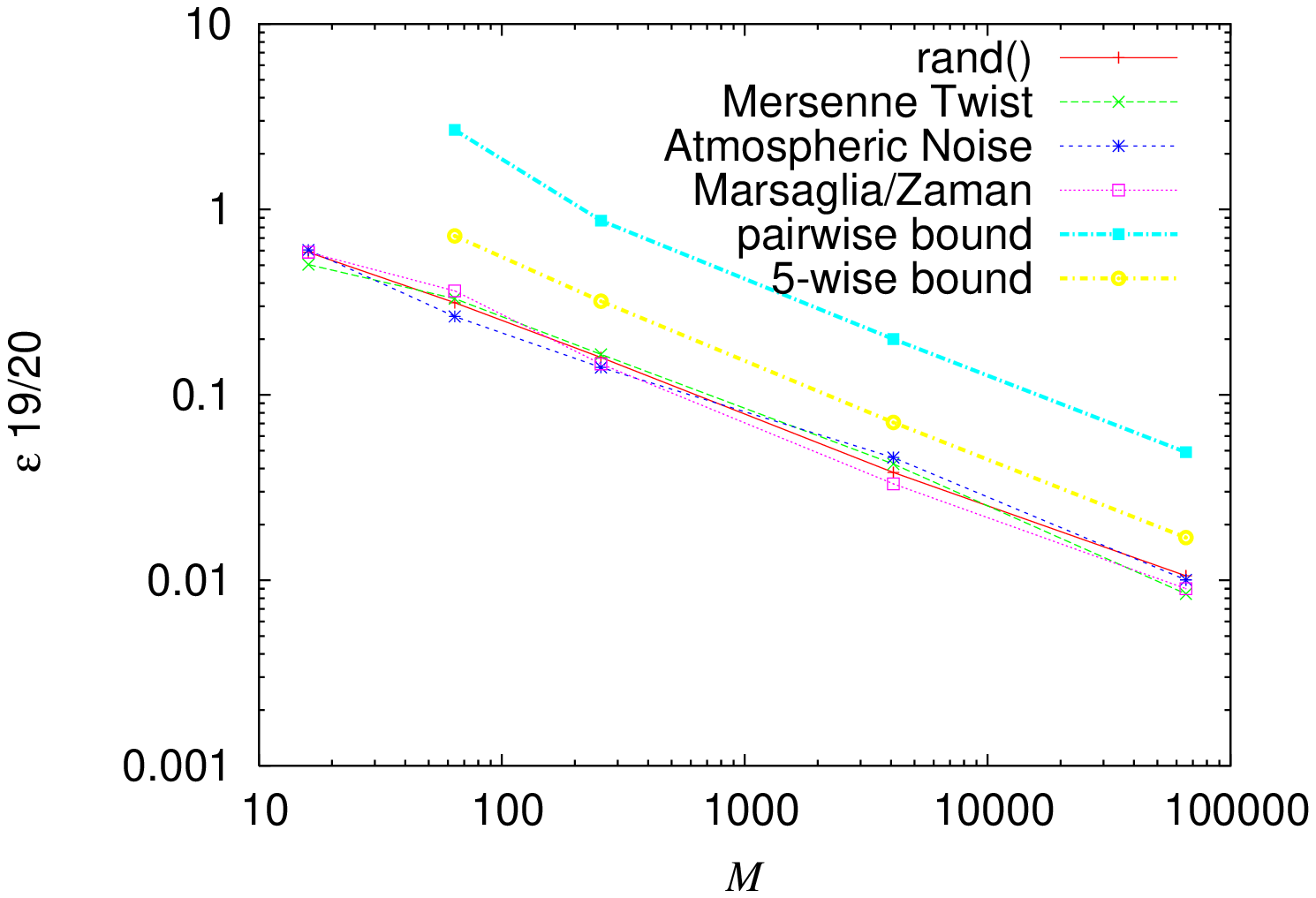}
}
\caption{The $n$-wise independent hash
is similarly unaffected}
\end{figure}

\end{document}